\documentclass[aps,pre,twocolumn,superscriptaddress,tight]{revtex4-1}

\usepackage{amssymb}
\usepackage{color}
\usepackage{graphicx}
\usepackage{amsmath}
\usepackage{mathrsfs}
\usepackage{times}
\usepackage{subeqnarray}
\usepackage{cases}
\usepackage{bm}
\usepackage{float}

\begin{document}

\title{Anticipating synchronization with machine learning}

\author{Huawei Fan}
\affiliation{School of Physics and Information Technology, Shaanxi Normal University, Xi'an 710062, China}

\author{Ling-Wei Kong}
\affiliation{School of Electrical, Computer, and Energy Engineering, Arizona State University, Tempe, Arizona 85287, USA}

\author{Ying-Cheng Lai}
\affiliation{School of Electrical, Computer, and Energy Engineering, Arizona State University, Tempe, Arizona 85287, USA}

\author{Xingang Wang}
\email{Email address: wangxg@snnu.edu.cn}
\affiliation{School of Physics and Information Technology, Shaanxi Normal University, Xi'an 710062, China}

\begin{abstract}

In applications of dynamical systems, situations can arise where it is desired to predict the onset of synchronization as it can lead to characteristic and significant changes in the system performance and behaviors, for better or worse. In experimental and real settings, the system equations are often unknown, raising the need to develop a prediction framework that is model free and fully data driven. We contemplate that this challenging problem can be addressed with machine learning. In particular, exploiting reservoir computing or echo state networks, we devise a ``parameter-aware'' scheme to train the neural machine using asynchronous time series, i.e., in the parameter regime prior to the onset of synchronization. A properly trained machine will possess the power to predict the synchronization transition in that, with a given amount of parameter drift, whether the system would remain asynchronous or exhibit synchronous dynamics can be accurately anticipated. We demonstrate the machine-learning based framework using representative chaotic models and small network systems that exhibit continuous (second-order) or abrupt (first-order) transitions. A remarkable feature is that, for a network system exhibiting an explosive (first-order) transition and a hysteresis loop in synchronization, the machine learning scheme is capable of accurately predicting these features, including the precise locations of the transition points associated with the forward and backward transition paths.

\end{abstract}
\date{\today}

\maketitle

\section{Introduction} \label{sec:intro}

As a universal phenomenon in nonlinear and complex dynamical systems, 
synchronization has attracted a great deal of research and a continuous 
interest~\cite{Kuramoto:book,PRK:book,Strogatz:book}. Synchronization 
represents a kind of coherent motion that typically arises in systems of 
coupled dynamical units when the interaction or coupling among the units is 
sufficiently strong. Depending on the specific form of the coherent motion, 
different types of synchronization can emerge, including complete chaotic 
synchronization~\cite{PC:1990}, phase synchronization~\cite{RPK:1996}, and 
generalized synchronization~\cite{KP:1996}. The occurrence of synchronization 
has significant consequences for the system behavior and functions. An 
example is the occurrence of epileptic seizures in the brain neural system. 
Specifically, a widely adopted assumption is that hypersynchrony is
closely associated with the occurrence of epileptic seizures~\cite{KSJ:book},
during which the number of independent degrees of freedom of the underlying 
brain dynamical system is reduced. Among the extensive literature in this
field, there was demonstration that partial and transient phase synchrony can 
be exploited to detect and characterize (but not to predict) seizure from 
multichannel brain data~\cite{LFO:2006,LFOH:2007,OL:2011}. To date, reliable
seizure prediction remains an unsolved problem. In real-world situations such 
as this, it is of interest to predict or anticipate synchronization before its 
actual occurrence. A daunting challenge is that it is often impossible to know 
the system equations so any prediction attempt must be based on time series 
data obtained before the system evolves into some kind of synchronous dynamical
state. We are thus motivated to ask the question: given that the system 
operates in a parameter regime where there is no synchronization, would it be 
possible to predict, without relying on any model, the onset of synchronization
based solely on the dynamically incoherent time series measurements taken from 
the parameter regime of desynchronization? In this paper, we articulate a 
machine-learning framework based on reservoir computing~\cite{Jaeger:2001,
JH:2004} to provide an affirmative answer to this question.

To place our work in a proper context, here we offer a brief literature 
review of the fields of synchronization and reservoir computing.

\paragraph*{Synchronization in nonlinear dynamics and complex networks.} 
In the study of synchronization, the typical setting is coupled dynamical 
oscillators, where the bifurcation or control parameter is the coupling 
strength among the oscillators. A central task is to identify the critical 
point at which a transition from desynchronization to synchronization 
occurs~\cite{Kuramoto:book,PRK:book}. Depending on the dynamics of the 
oscillators and the coupling function, the system can have a sequence of 
transitions, giving rise to distinct synchronization regimes in the parameter 
space. For example, for a system of coupled identical nonlinear oscillators, 
complete synchronization can arise when the coupling exceeds a critical 
strength that can be determined by the master stability 
function~\cite{PC:1998,HCLP:2009}. Systems of nonlinearly coupled phase 
oscillators, e.g., those described by the classic Kuramoto 
model~\cite{Kuramoto:book} can host phase synchronization and the critical 
coupling strength required for the onset of this type of ``weak'' 
synchronization can be determined by the mean-field theory~\cite{WS:1993,
OA:2008}. Synchronization in coupled physical oscillators was experimentally
studied~\cite{WMRSDS:2013,WSMR:2013}. A counter-intuitive phenomenon is 
that adding connections can hinder network synchronization of time-delayed 
oscillators~\cite{HPPMR:2015}. With the development of modern network science 
in the past two decades, synchronization in complex networks has been 
extensively studied~\cite{ADKMZ:2008}, due to the rich variety of complex 
network structures in natural and engineering systems and the fundamental role 
of synchronous dynamics in the system functioning. Earlier it was found that 
small-world networks, due to their small network diameter values, are more 
synchronizable than regular networks of comparable sizes~\cite{BP:2002}, but 
heterogeneity in the network structure presents an obstacle to 
synchronization-\cite{NMLH:2003}. Subsequently it was found that heterogeneous
networks with weighted links can be more synchronizable than small-world and 
random networks~\cite{WLL:2007}. The onset of chaotic phase synchronization 
in complex networks of coupled heterogeneous oscillators was also 
studied~\cite{RTHL:2012}. The interplay between network symmetry and 
synchronization was uncovered and understood~\cite{PSHMR:2014,SPHMR:2016}.
In complex networks, the transition from desynchronization to synchronization 
is typically continuous, e.g., the synchronization error or the order 
parameter tends to change continuously through the critical point, which is 
characteristic of a second-order phase transition. However, studies of 
networked systems of coupled phase oscillators revealed that, if the network 
links are weighted according to the natural frequencies of the oscillators, 
the transition can be abrupt and discontinuous, i.e., a first-order phase 
transition~\cite{GGAM:2011,ZPSLK:2014,BAGLLSWZ:2016}. This phenomenon, known 
as explosive synchronization, is proof of the strong influence of network 
structure on the collective dynamics. 

\paragraph*{Reservoir computing for predicting chaotic systems.}
The idea and principle of exploiting reservoir computing for predicting the
state evolution of chaotic systems were first laid out about two decades
ago~\cite{Jaeger:2001,JH:2004}. In recent years, model-free predication of 
chaotic systems using reservoir computing has gained considerable 
momentum~\cite{LBE:2015,HSRFG:2015,LBMUCJ:2017,PLHGO:2017,LPHGBO:2017,DBN:book,LHO:2018,PWFCHGO:2018,PHGLO:2018,Carroll:2018,NS:2018,ZP:2018,WYGZS:2019,GPG:2019,JL:2019,VPHSGOK:2019,FJZWL:2020,ZJQL:2020,GYL:2021}. 
The neural architecture of reservoir computing consists of a single 
hidden layer of a complex dynamical network that receives input data and 
generates output data. In the training phase, the whole system is set to the 
``open-loop'' mode, where it receives input data and optimizes its parameters 
to match its output with the true output corresponding to the input. In the 
standard reservoir computing setting~\cite{LBE:2015,HSRFG:2015,LBMUCJ:2017,PLHGO:2017,LPHGBO:2017,DBN:book,LHO:2018,PWFCHGO:2018,PHGLO:2018,Carroll:2018,NS:2018,ZP:2018,WYGZS:2019,GPG:2019,JL:2019,VPHSGOK:2019,FJZWL:2020,ZJQL:2020,GYL:2021},
the adjustable parameters are those associated with the output matrix that 
maps the internal dynamical state of the hidden layer to the output layer, 
while other parameters, such as those defining the network and the input 
matrix, are fixed (the hyperparameters). In the prediction phase, the neural 
machine operates in the ``close-loop'' mode where the output variables are 
fed directly into the input, so that the whole system becomes a self-evolving 
dynamical system. With reservoir computing, the state evolution of chaotic 
systems can be accurately predicted for about half dozen Lyapunov time - 
longer than that can be achieved using the traditional methods of nonlinear 
time series analysis. A result was that, even reservoir computing is unable to 
make long term prediction of the state evolution of a chaotic system, it is 
still able to replicate the ergodic properties of the system~\cite{PLHGO:2017}. 
This feature makes it possible to generate the bifurcation diagram of a 
nonlinear dynamical system without the equations~\cite{CA:2019}. More recently,
a ``parameter-aware'' reservoir computing scheme was articulated to predict 
critical transitions, transient chaos, and system collapse~\cite{KFGL:2021}. 

\paragraph*{Contributions of the present work.}
In situations where synchronization is undesired, such as epilepsy, the sudden
onset of synchronization, or a ``synchronization catastrophe,'' is of concern.
Likewise, in alternative situations where synchronization is deemed desired,
a sudden collapse of synchronization is to be avoided. It is of interest to
predict sudden onset or collapse of synchronization before it occurs. If the
detailed network structure and system equations are known, in principle the
prediction task is feasible. In real world applications, it is often the case
that the only available information about the system is a set of measured time
series at a number of control parameter values~\cite{RPK:1996,KP:1996}. For
instance, in epilepsy, the details of the brain neuronal network generating
the seizures are unknown, but the EEG signals upon applications of controlled
drug doses can be obtained. In such applications, a model-free and fully
data-driven approach to predicting a synchronization catastrophe is needed.
In this regard, if the network is sparse and if the nodal dynamical equations
have a simple mathematical structure, such as those which can be represented
by a small number of power series or Fourier series terms, sparse optimization
tailored to nonlinear dynamical systems~\cite{WYLKG:2011,WYLKH:2011,WLG:2016}
can be employed to discover the system equations so as to predict the onset
of synchronization~\cite{SNWL:2012}. 

The scheme of reservoir computing incorporating a parameter input channel so 
that it is able to sense and track the parameter variations of the target 
system~\cite{KFGL:2021} provides a solution to the problem of predicting 
synchronization transition. The purpose of this paper is to demonstrate that 
this is indeed the case. The basic principle is that the dynamical ``climate'' 
of the target system, e.g., the dimension of its chaotic 
attractor~\cite{PLHGO:2017}, undergoes an abrupt change at the synchronization 
transition point, and a well trained parameter-aware reservoir computing 
machine is able to capture this ``climate'' change. To be concrete, we take 
the coupling strength as the control parameter whose value is fed into the 
input parameter channel. The required training constitutes optimizing the
reservoir output matrix based on time series collected from a small number of
distinct values of the control parameter in the desynchronization regime so 
that the neural machines learns the ``climate change'' of the target system. 
As we will show, provided that the training is successful, the machine is able 
to predict a characteristic change in the collective dynamical behavior of the 
target system for any value of the control parameter that the input parameter 
channel receives. The scenarios tested consist of a broad spectrum of complex 
synchronization behaviors, including complete synchronization in coupled 
identical chaotic systems and explosive synchronization in networks of coupled 
nonidentical phase oscillators. Because of the model-free nature of our 
machine-learning framework for synchronization prediction, it can be applied 
directly to real world systems.

\section{Machine-learning method} \label{sec:rc}

Our reservoir computing machine consists of four modules: the $I/R$ layer 
(input-to-reservoir), the control module, the hidden layer (the reservoir), 
and the $R/O$ layer (reservoir-to-output). The $I/R$ layer is characterized 
by $\mathcal{W}_{in}$, a $D_r\times D_{in}$-dimensional matrix that maps the 
input vector $\mathbf{u}_{\varepsilon}(t)\in \mathbf{R}^{D_{in}}$ to the 
dynamical network in the reservoir hidden layer, where the input vector is 
acquired from the target system at time $t$ for the specific bifurcation 
parameter value $\varepsilon$. The elements of $\mathcal{W}_{in}$ are randomly 
drawn from a uniform distribution within the range $[-\sigma,\sigma]$. The 
control module is characterized by the vector $\mathbf{s}=\eta(t)\mathbf{b}$, 
where $\eta(t)$ is the time-dependent control parameter and 
$\mathbf{b}\in \mathbb{R}^{D_{r}}$ is a bias vector. In general, the control 
parameter is related to the bifurcation parameter of the target system by a 
smooth function, where a convenient choice is simply $\eta(t)=\varepsilon(t)$. 
Effectively, $\eta(t)$ can be regarded as an additional input component that 
can be incorporated into the input vector $\mathbf{u}(t)$, where the elements 
of $\mathbf{b}$ are also drawn randomly from a uniform distribution within the 
range $[-\sigma,\sigma]$. The network in the hidden layer consists of $D_r$ 
nonlinear elements (nodes), whose dynamics are governed by the rule 
\begin{eqnarray} \label{eq:rc1}
\mathbf{r}(t+\Delta t)&=&(1-\alpha)\mathbf{r}(t) +\alpha\tanh[\mathcal{A}\mathbf{r}(t)\notag \\
        & &+\mathcal{W}_{in}\mathbf{u}_{\varepsilon}(t)+\varepsilon_k(\varepsilon(t)+\varepsilon_{bias})\mathbf{b}],
\end{eqnarray}
where $\mathbf{r}(t)\in \mathbb{R}^{D_r}$ is the state vector of the network 
at time $t$, $\Delta t$ is the time step, $\alpha$ is a leakage parameter,
$\varepsilon_k$ and $\varepsilon_{bias}$ define a linear transformation of
$\varepsilon$ before input it into the reservoir, and $\mathcal{A}$ is the 
$D_r\times D_r$-dimensional matrix characterizing the connecting structure of 
the hidden layer network. With probability $p$, the elements of matrix 
$\mathcal{A}$ are set to be zero. The symmetric, non-zero elements of 
$\mathcal{A}$ are drawn from a uniform distribution within the range $[-1,1]$, 
and are normalized so as to make the spectral radius of the matrix $\rho$. The 
output layer is characterized by a $D_{out}\times D_{r}$ dimensional matrix 
$\mathcal{W}_{out}$, which generates the $D_{out}$-dimensional output vector 
$\mathbf{v}(t)$ via 
\begin{equation} \label{eq:rc2}
\mathbf{v}(t+\Delta t)=\mathcal{W}_{out}\cdot\mathbf{f}(\mathbf{r}(t+\Delta t)),
\end{equation}
where $\mathbf{f}(\mathbf{r})$ is the output function and we set 
$D_{in}=D_{out}$. The elements of the output matrix are to be determined 
through training, where a general rule is to set 
$\mathbf{f}(\mathbf{r}) = \mathbf{r}$ for the odd nodes in the reservoir and 
$\mathbf{f}(\mathbf{r})=\mathbf{r}^2$ for the even nodes so as to enable 
proper optimization~\cite{PHGLO:2018}. In particular, different from 
$\mathcal{W}_{in}$, the elements of $\mathcal{W}_{out}$ are not known a priori,
but are to be ``learned'' from the input data through training, with the 
purpose to find the proper matrix $\mathcal{W}_{out}$ such that the output 
vector $\mathbf{v}(t+\Delta t)$ as calculated from Eq.~(\ref{eq:rc2}) is as 
close as possible to the input vector $\mathbf{u}(t+\Delta t)$ for 
$t=(\tau+1)\Delta t,\ldots,(L-1)\Delta t,L\Delta t$, where $T_0=\tau\Delta t$ 
is the initial period discarded to remove the transient behavior in reservoir's
response to the training signal, and $L$ is the length of the training time 
series. This can be done~\cite{PLHGO:2017,LPHGBO:2017,PHGLO:2018} by 
minimizing a cost function with respect to $\mathcal{W}_{out}$, which gives
\begin{equation} \label{eq:rc3}
\mathcal{W}_{out}=\mathcal{U}\mathcal{F}^T(\mathcal{F}\mathcal{F}^T+\lambda \mathcal{I})^{-1},
\end{equation}
where $\mathcal{F}$ is the $D_{r}\times L$ dimensional state matrix whose $k$th
column is $\mathbf{f}(\mathbf{r}((\tau+k)\Delta t))$, $\mathcal{U}$ is the 
$D_{in}\times L$ dimensional matrix whose $k$th column is 
$\mathbf{u}((\tau+k+1)\Delta t)$, $\mathcal{I}$ is the identity matrix, and 
$\lambda$ is the ridge regression parameter. 

After training, the elements in matrix $\mathcal{W}_{out}$ are fixed, and the
machine is ready for prediction, where we first set the control parameter to 
a specific value of interest (not necessarily any of the parameter values used 
in the training phase), and then evolve the machine according to 
Eq.~(\ref{eq:rc1}) by replacing $\mathbf{u}_{\varepsilon}(t)$ with 
$\mathbf{v}(t+\Delta t)$. Finally, by tuning $\varepsilon$ to different 
values, we monitor the variation of the statistical properties of the 
reservoir outputs, and predict the transition of the system dynamics with 
respect to $\varepsilon$.

The main feature of our reservoir computing design is that the input data in 
the training phase contain two components: (1) the input vector 
$\mathbf{u}_{\varepsilon}(t)$ representing the time series measured from the 
target system and (2) the bifurcation parameter $\varepsilon (t)$ under which 
$\mathbf{u}_{\varepsilon}(t)$ is obtained, whereas in the conventional
scheme~\cite{LBE:2015,HSRFG:2015,LBMUCJ:2017,PLHGO:2017,LPHGBO:2017,DBN:book,LHO:2018,PWFCHGO:2018,PHGLO:2018,Carroll:2018,NS:2018,ZP:2018,WYGZS:2019,GPG:2019,JL:2019,VPHSGOK:2019,FJZWL:2020,ZJQL:2020,GYL:2021}, only the first component
(time series from a fixed value of the bifurcation parameter) is present.
In particular, $\mathbf{u}_{\varepsilon}(t)$ consists of $m$ segments of equal 
length $T$ (i.e., $L=mT$) and, for each segment, the value of the bifurcation 
parameter $\varepsilon (t)$ is fixed so, overall, $\varepsilon(t)$ is a step 
function of time. (The proposed scheme is equally effective when the segments 
are not of equal length - see Appendix~\ref{appendix_a}.) In the predicating 
phase, the input vector $\mathbf{u}_{\varepsilon}(t)$ is replaced by 
$\mathbf{v}(t)$ as in the conventional scheme, but the value of the bifurcation
parameter $\varepsilon (t)$ is still needed as an input. Since our goal is to 
predict synchronization among a number of coupled oscillators, the coupling 
strength $\varepsilon$ is a natural choice for the bifurcation parameter. 

\section{Results}

\subsection{Predicting complete synchronization in coupled chaotic maps} 
\label{subsec:logisticmap}

\begin{figure*}[ht!]
\centering
\includegraphics[width=\linewidth]{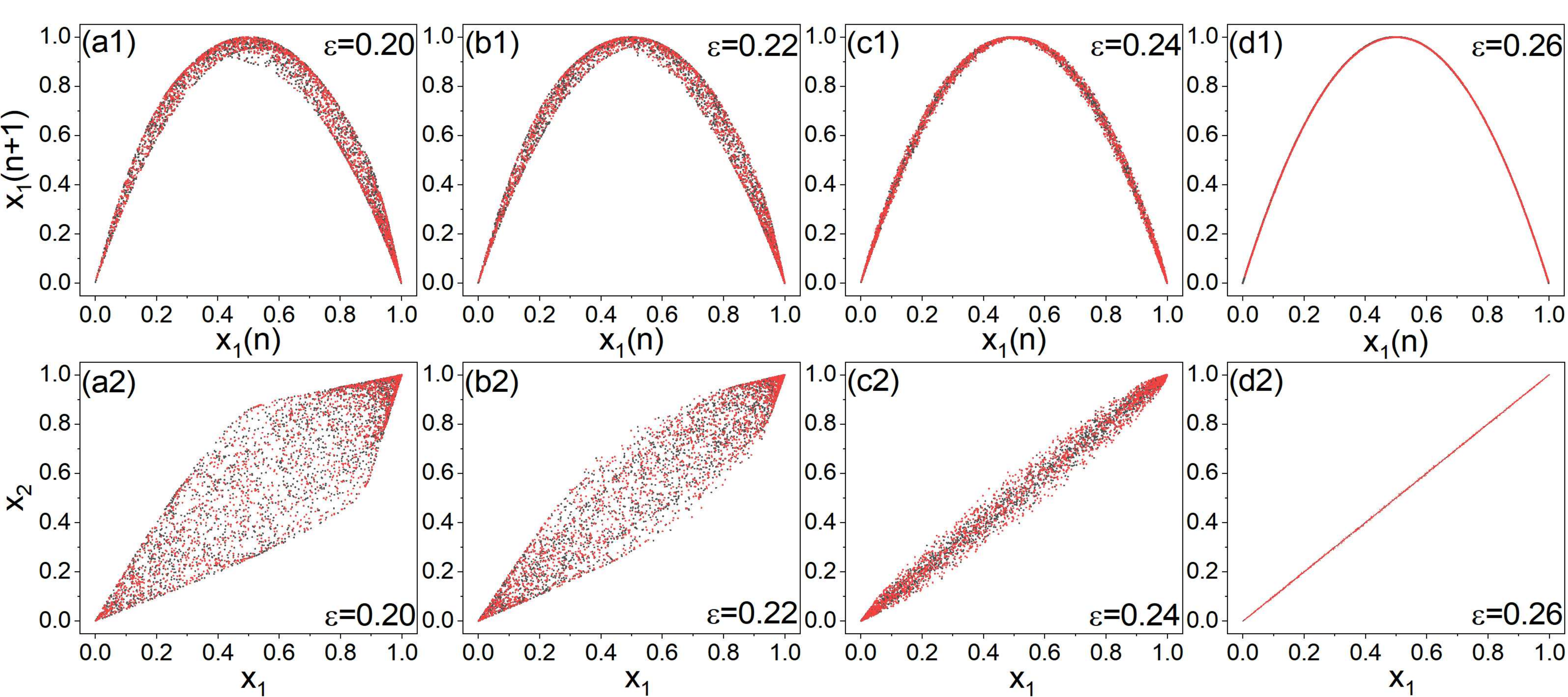}
\caption{Predicted synchronization behavior for different values of the
bifurcation parameter in coupled chaotic maps. The system consists of a pair 
of coupled chaotic logistic maps with coupling parameter $\varepsilon$ and
respective dynamical variables $x_1$ and $x_2$. Top row (a1-d1): the predicted
(black dots) and true (red dots) returned map constructed from $x_1$ for 
$\varepsilon=0.2$, 0.22, 0.24, and 0.26, respectively; bottom row (a2-d2): 
the predicted (black) and true (red) mutual relationship between $x_1$ and 
$x_2$ for the same set of parameter values, where a diagonal line represents 
complete synchronization.} 
\label{fig:LGM}
\end{figure*}

\begin{figure}[ht!]
\centering
\includegraphics[width=0.93\linewidth]{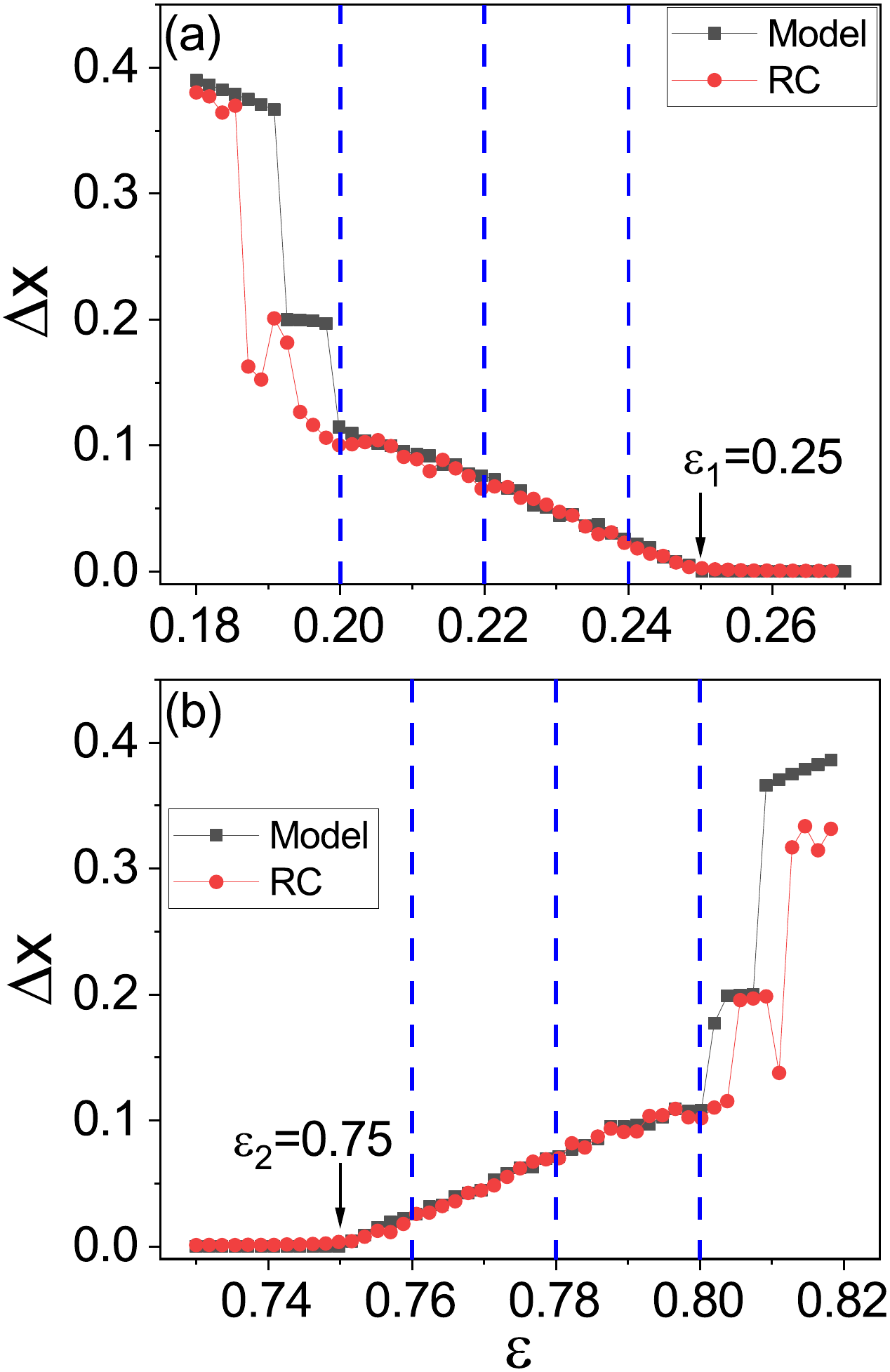}
\caption{Predicting synchronization transition in coupled chaotic logistic 
maps. (a) As the coupling is strengthened, the synchronization error $\Delta x$ 
gradually decreases to zero at about $\varepsilon_1 \approx 0.25$. (b) The
error $\Delta x$ starts to increase from zero at about 
$\varepsilon_2 \approx 0.75$. The vertical dashed lines denote the coupling
parameter values used in generating the training data. The machine-predicted
and true results are represented as red circles and black squares, 
respectively. The machine predicts correctly the transitions at both ends 
of the synchronization parameter regime $(\varepsilon_1,\varepsilon_2)$.}
\label{fig:LGM_2}
\end{figure}

We consider the following system of two coupled, identical chaotic maps:
\begin{equation} \label{coupledlog}
\mathbf{x}_{1,2}(n+1)=\mathbf{F}(\mathbf{x}_{1,2}(n))+\varepsilon[\mathbf{H}(\mathbf{x}_{2,1}(n))-\mathbf{H}(\mathbf{x}_{1,2}(n))],
\end{equation}
where $\mathbf{x}_{1,2}(n)$ denote the dynamical variables of the system at 
the $n$th iteration, $\mathbf{F}(\mathbf{x})$ and $\mathbf{H}(\mathbf{x})$ are
the map and coupling functions, respectively. As an illustrative example,
we choose the one-dimensional chaotic logistic map defined on the unit 
interval: $F(x)=4x(1-x)$, and set the coupling function to be $H(x)=F(x)$. 
The critical coupling value for complete synchronization can be obtained 
using the master stability function~\cite{PC:1998,HCLP:2009}, which gives
that complete synchronization occurs for 
$0.25 \approx \varepsilon_1 \le \varepsilon \le \varepsilon_2 \approx 0.75$. 

We obtain the training data for three different values of $\varepsilon$: 
0.2, 0.22 and 0.24, all in the desynchronization regime.
For each $\varepsilon$ value, we collect the state variables 
$\{\mathbf{u}(n)\}=\{x_1 (n),x_2(n)\}$ for $T=2\times 10^3$ successive time 
steps after disregarding a transient of $10^3$ time steps. The time series 
from the three values of $\varepsilon$ are combined to form a single time 
series which, together with the step function $\varepsilon(n)$ of the coupling
parameter, are fed into the reservoir for training that yields the optimal
output matrix $\mathcal{W}_{out}$. The values of the hyperparameters of the 
reservoir are $(D_r,p,\sigma,\rho,\alpha,\varepsilon_k,\varepsilon_{bias})
=(100,0.2,1,10^{-5},1,1,0)$. The regression 
parameter for obtaining $\mathcal{W}_{out}$ is $\lambda = 1\times 10^{-5}$.

To predict the synchronization transition, the trained reservoir computing 
machine must possess the ability to sense the change in the ``synchronization 
climate'' of the target system. Figures~\ref{fig:LGM}(a1-d1) show the 
predicted return maps (black dots) constructed from $x_1$ for four values of 
$\varepsilon$: 0.2, 0.22, 0.24, and 0.26, respectively, together with the true 
return maps (red dots). The corresponding plots of the mutual relation between 
the two maps are shown in Figs.~\ref{fig:LGM}(a2-d2). The first three values 
of $\varepsilon$ are below $\varepsilon_1$, so there is no synchronization, 
and the last value is in the synchronization regime. The reservoir machine 
predicts these behaviors correctly. Especially, as the value of $\varepsilon$ 
is increased from 0.2 to 0.26, the return map gradually evolves into the map 
function $F(x)=4x(1-x)$ and the points $(x_1,x_2)$ converges to the diagonal 
line, which are characteristic of complete synchronization. In fact, 
statistically the black and red dots cannot be distinguished, indicating the 
superior power of the machine to capture the collective dynamics of the target 
system. 

Note that the first three $\varepsilon$ values (0.2, 0.22, and 0.24) are the 
ones used in training. It may thus not be surprising that the reservoir is 
able to predict correctly the distinct dynamical behaviors of the system at 
these parameter values, i.e., there is no synchronization. What is remarkable 
is that the last $\varepsilon$ value (0.26) is totally ``new'' to the machine 
as it has never been exposed to data from this parameter value, yet it 
predicts, still quite correctly, that there is now synchronization. This means 
that training at different coupling parameter values in the desynchronization
regime has instilled into the machine the ability for it to ``sense'' the 
``climate'' change in the collective dynamics of the target system.

As the reservoir computing has been trained to capture the ``climate'' of 
the collective dynamics in the coupled chaotic map system, it should be able
to predict the synchronization transition. In particular, the expectation is
that it would predict correctly the two ending points of the synchronization 
parameter regime $(\varepsilon_1,\varepsilon_2)$ at which a transition to 
synchronization occurs depending on the direction of parameter variation. 
To demonstrate the successful prediction of the transition point 
$\varepsilon_1$, we fix the output matrix $\mathcal{W}_{out}$ and increase the 
$\varepsilon$ value systematically from $0.18$ to $0.27$ at the step size 
$\Delta \varepsilon=1\times 10^{-3}$. For each $\varepsilon$ value, we let the 
machine generate a time series of length $T=1\times 10^3$ and calculate the 
time-averaged synchronization error $\Delta x=\left<|x_1-x_2|\right>_T$. 
Figure~\ref{fig:LGM_2}(a) shows the machine-predicted $\Delta x$ versus 
$\varepsilon$ (red circles), where $\Delta x$ decreases continuously for 
$\varepsilon \agt 0.2$ and becomes zero for 
$\varepsilon = \varepsilon_1\approx 0.25$. For comparison, the true 
behavior of $\Delta x$ is also included in Fig.~\ref{fig:LGM_2}(a), to which
the predicted result agrees well for $\varepsilon \agt 0.2$. At the opposite
end of the synchronization regime, the reservoir machine does an equally 
good job to predict the transition to synchronization at 
$\varepsilon_1\approx 0.75$ as $\varepsilon$ decreases from a larger value,
as shown in Fig.~\ref{fig:LGM_2}(b), where the training data are obtained 
from $\varepsilon=0.76$, $0.78$ and $0.8$. The results in Fig.~\ref{fig:LGM_2} 
are thus evidence that a properly trained reservoir computing machine has the 
power to accurately predict the critical point of transition to 
synchronization. 

The performance of the reservoir machine in predicting the synchronization
transition is affected slightly by the number and locations of the training 
parameter values. We find that, insofar as training is done with data from at 
least two distinct parameter values, the synchronization transition can be
predicted. Training with data from increasingly more values of the bifurcation
parameter, regardless of the order, can in general improve the prediction 
accuracy. Also, the closer the training parameter values to the critical 
point, the more accurate the prediction. (Details about the effects of the 
number, locations and order of the training parameter values on the prediction 
performance are provided in Appendix~\ref{appendix_a}.)

\subsection{Predicting synchronization transition in coupled chaotic Lorenz 
oscillators} \label{subsec:Lorenz}

The system of a pair of coupled identical chaotic Lorenz oscillators is given 
by
\begin{align}\label{lorenz}
&\dot{x}_{1,2}=\mu (y_{1,2}-x_{1,2})+\varepsilon(x_{2,1}-x_{1,2}) \notag, \\
&\dot{y}_{1,2}=x_{1,2}(\beta-z_{1,2})-y_{1,2}+\varepsilon(y_{2,1}-y_{1,2}), \\
&\dot{z}_{1,2}=x_{1,2}y_{1,2}-\gamma z_{1,2}+\varepsilon(z_{2,1}-z_{1,2})\notag,
\end{align}
where the parameter setting is $(\mu,\beta,\gamma)=(10,28,2)$ for which an 
isolated oscillator has a chaotic attractor~\cite{Lorenz:1963}. Analysis based 
on the master stability function gives that complete synchronization occurs 
for $\varepsilon > \varepsilon_c \approx 0.42$. 

\begin{figure}[tbp]
\centering
\includegraphics[width=\linewidth]{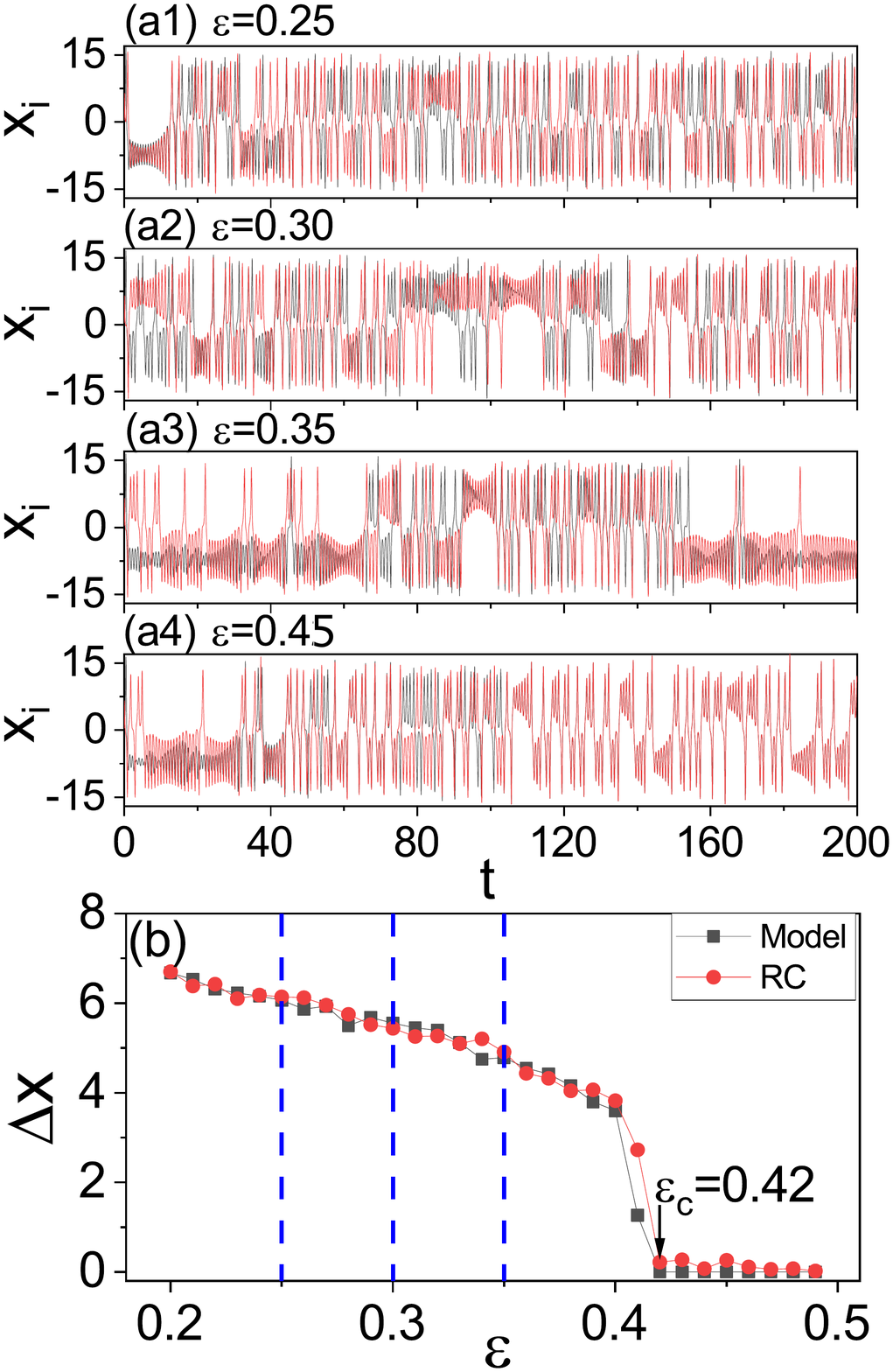}
\caption{Predicting synchronization transition in coupled chaotic Lorenz 
oscillators. (a1-a3) Machine-generated time evolution of $x_{1}$ (black trace) 
and $x_{2}$ (red trace) for the three training values of the control parameter 
in the desynchronization regime: $\varepsilon = 0.25$, 0.3, and 0.35. 
(a4) Predicted synchronization behavior for $\varepsilon = 0.45$. The machine 
has never been exposed to data from this parameter value, yet it successfully 
predicts synchronization. (b) Synchronization error $\Delta x$ versus 
$\varepsilon$. Red circles and black squares represent the machine-predicted 
and true results, respectively. The three vertical dashed lines indicate the 
locations of the three training parameter values.}
\label{fig:Lorenz}
\end{figure}

We generate the training data from three distinct values of the coupling
parameter in the desynchronization regime: $\varepsilon=0.25$, $0.3$ and 
$0.35$. For each parameter value, a time series of length $T=4\times 10^3$ 
is collected at the time step $\Delta t=0.02$, after disregarding a transient
phase of length $T_0=5\times 10^3$. The input vector 
$\mathbf{u}(t) \equiv [x_1(t),y_1(t),z_1(t),x_2(t),y_2(t),z_2(t)]^T$ and the 
control parameter signal $\varepsilon(t)$ are fed into the reservoir 
machine for training the output matrix $\mathcal{W}_{out}$. The hyperparameters
of the reservoir are set as $(D_r,p,\sigma,\rho,\alpha,\varepsilon_k,
\varepsilon_{bias})=(2\times 10^3,0.2,0.05,0.1,1,1,0)$, 
and the regression parameter value is $\lambda=1\times 10^{-8}$. 

Figures~\ref{fig:Lorenz}(a1-a3) show the predicted evolution of the dynamical 
variables $x_1$ and $x_2$ from the two oscillators for the three training 
parameter values preceding the onset of synchronization. The machine predicts 
correctly that there is no synchronization for these parameter values. When 
the coupling parameter value is set to be $\varepsilon=0.45$ (in the 
synchronization regime), the machine indeed predicts the synchronous behavior, 
as shown in Fig.~\ref{fig:Lorenz}(a4). To test if the machine can predict the 
critical transition point $\varepsilon_c$ for synchronization, we increase the 
control parameter value in the machine from $0.2$ to $0.5$ systematically and 
calculate the time-averaged synchronization error 
$\Delta x=\left<|x_1-x_2|\right>_T$ over a time period of $T=1\times 10^4$ 
(after discarding transients). Figure~\ref{fig:Lorenz}(b) shows $\Delta x$ 
versus $\varepsilon$. The machine predicted that the synchronization error 
approaches zero for $\varepsilon_c \approx 0.42$, which is in good agreement 
with the true value of the transition point. Similar results are obtained for 
alternative coupling configurations, e.g., through a single variable or a 
cross coupling scheme (Appendix~\ref{appendix_b}).

\subsection{Predicting synchronization transition in coupled chaotic food-chain systems} \label{subsec:fc}

We consider the following mutually coupled system of two food chains, each 
with three-species~\cite{MY:1994}:
\begin{eqnarray}
\dot{x}_{1,2} & = & x_{1,2}(1-\frac{x_{1,2}}{K})-\frac{a_c b_c x_{1,2} y_{1,2}}{x_{1,2}+x_0}, \\ 
\dot{y}_{1,2} & = & a_c y_{1,2}[\frac{b_c x_{1,2}}{x_{1,2}+x_0}-1]-\frac{a_p b_p y_{1,2} z_{1,2}}{y_{1,2}+y_0}\notag \\
& & + \varepsilon (y_{2,1}-y_{1,2}), \\
\dot{z}_{1,2} & = & a_p z_{1,2}(\frac{b_p y_{1,2}}{y_{1,2}+y_0}-1) + \varepsilon (z_{2,1}-z_{1,2}),
\end{eqnarray}
where $x_{1,2}$, $y_{1,2}$, $z_{1,2}$ are the population densities of the 
resource, consumer, and predator species, respectively, in each food chain.
We use the parameter setting~\cite{MY:1994} by which each isolated food chain 
is a chaotic oscillator:
$(K,a_c,b_c,a_p,b_p,x_0,y_0)=(0.99,0.4,2.009,0.08,2.876,0.16129,0.5)$.
Under this setting, complete synchronization occurs for 
$\varepsilon > \varepsilon_c \approx 8.4\times10^{-3}$. 

We generate the training data from three distinct values of the coupling
parameter: $\varepsilon=4.5\times10^{-3}$, $5.5\times10^{-3}$, and
$6.5\times10^{-3}$, all in the desynchronization regime. For each value,
we generate a time series of length $T=9,600\delta t$ with $\delta t=0.5$, 
after disregarding the initial $T_0=8,000$ steps to remove any transient 
behavior. All six dynamical variables of the coupled system as well as the 
coupling strength are fed as input to the reservoir machine. The values of
the hyperparameters of the reservoir machine are set as 
$(D_r,p,\sigma,\rho,\alpha,\varepsilon_k,\varepsilon_{bias})=
(1000,0.695,2.30,1.20,0.27,0.049,1.53)$, where the elements of the reservoir
network $\mathcal{A}$ are sampled from a standard normal distribution, and
the regression parameter is $\lambda=3\times10^{-4}$.

\begin{figure}[ht!]
\centering
\includegraphics[width=\linewidth]{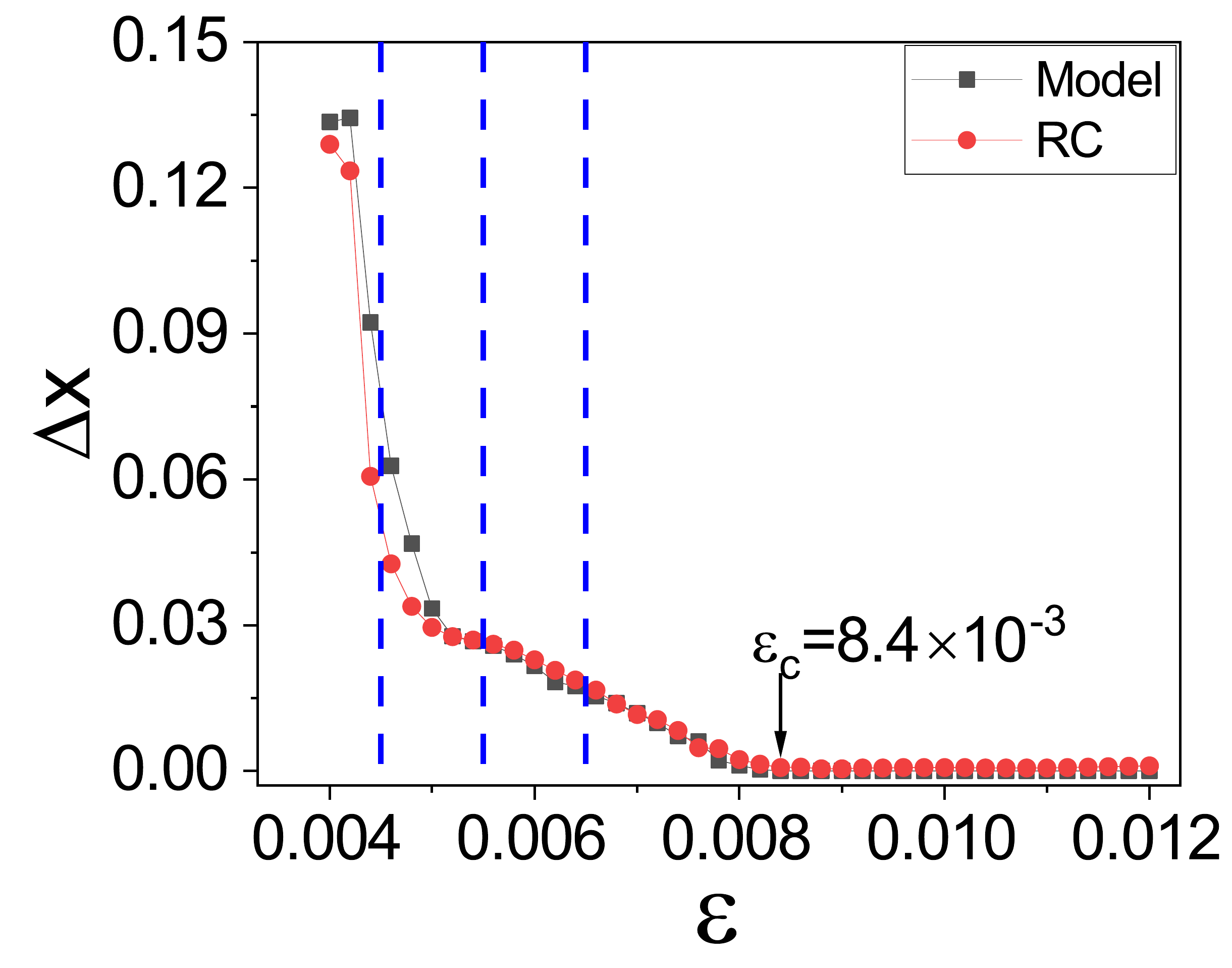}
\caption{Predicting synchronization transition in coupled chaotic food chains.
Shown are the true synchronization errors (black squares) and the reservoir’s 
predicted errors (red circles) for various values of the coupling strength 
$\varepsilon$. The three vertical blue dashed lines indicate the locations of 
the training parameter values.}
\label{fig:FoodChain}
\end{figure}

We use the trained reservoir to predict the synchronization behaviors in the 
coupling range $\varepsilon \in[0.04,0.012]$, by calculating the time-averaged 
synchronization error $\Delta x=\left<|x_1-x_2|\right>_T$ over a time period of
$T=2.4\times 10^4$ steps for each parameter value (after discarding transients 
with $T_0=4.8\times 10^4$ steps). Figure~\ref{fig:FoodChain} shows that the 
predicted error versus $\varepsilon$ (red circles) is quite close to the true 
errors (black squares). The machine predicted synchronization transition point 
agrees with the true value well. This is quite remarkable considering that 
only the time series from the desynchronization regime are used as the 
training data.

\subsection{Predicting explosive synchronization in coupled nonidentical phase oscillators} \label{subsec:explo}

The route to complete synchronization treated in Secs.~\ref{subsec:logisticmap},
\ref{subsec:Lorenz}, and \ref{subsec:fc} belongs to the category of 
second-order phase transition, where a physical quantity characterizing the 
degree of synchronization varies continuously (albeit non-smoothly) through 
the transition point. As demonstrated, our machine learning scheme is fully 
capable of predicting the transition in all the cases. Another type of 
synchronization transition that has been extensively studied is first-order 
phase transition, also known as explosive synchronization~\cite{GGAM:2011,
ZPSLK:2014,BAGLLSWZ:2016}, where the onset of synchronization is abrupt in the 
sense that the underlying characterizing quantity changes discontinuously at 
the transition point. The question is whether our machine learning scheme can 
predict explosive synchronization. Here we present an affirmative answer using 
the paradigmatic system in the literature~\cite{GGAM:2011,ZPSLK:2014,
BAGLLSWZ:2016} for explosive synchronization: coupled nonidentical phase 
oscillators. 

\begin{figure}[tbp]
\centering
\includegraphics[width=\linewidth]{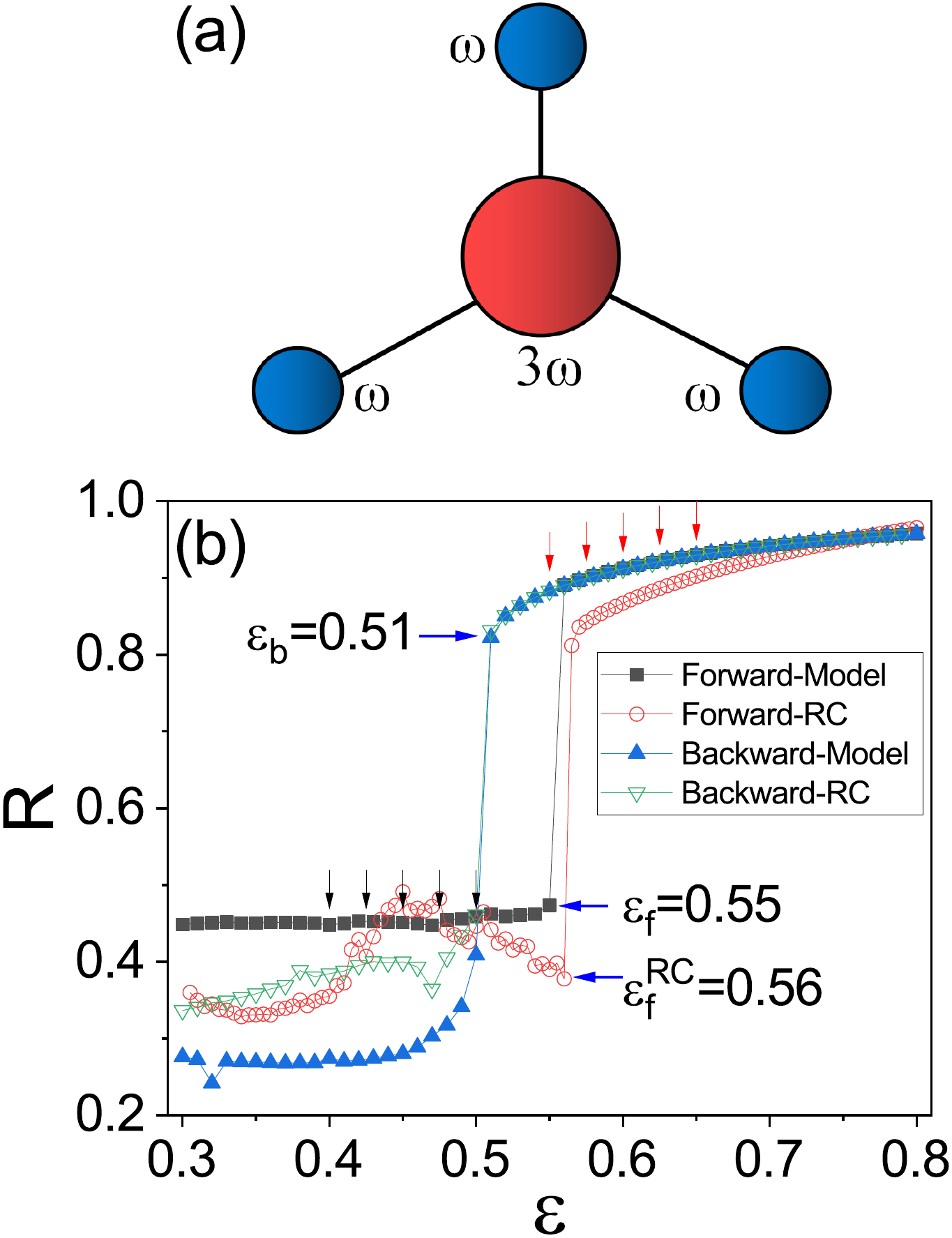}
\caption{ Predicting explosive synchronization transitions in coupled 
nonidentical phase oscillators. (a) A star network. (b) Synchronization order 
parameter $R$ versus the coupling strength $\varepsilon$ for the forward and 
backward transition paths. The true transitions obtained from model simulations
are represented by black solid squares (forward transition) and blue solid 
up-triangles (backward transition). The corresponding machine predictions are 
displayed as red open circles (forward transition) and green empty 
down-triangles (backward transition). The values of the coupling parameters 
used to generate the training data for predicting the forward and backward 
transitions are marked by the black and red arrows, respectively.}
\label{fig:explosive}
\end{figure}

For simplicity, we consider a small star network of $N=4$ nodes, as shown in
Fig.~\ref{fig:explosive}(a), where the natural frequencies of the peripheral 
(leaf) nodes are identical and that of the hub node is proportional to its 
degree~\cite{ZPSLK:2014}. The network dynamics is described by 
\begin{align} \label{eq:starnet}
&\dot{\theta}_{l}=\omega+\varepsilon\sin(\theta_{h}-\theta_{l})\notag, \\
&\dot{\theta}_{h}=k_h\omega+\varepsilon\sum_{l=1}^{3}\sin(\theta_{l}-\theta_{h}),
\end{align}
where $l=1,2,3$ denote the leaf nodes, $h$ denotes the hub node, $\varepsilon$ 
is the uniform coupling strength, and $k_h=3$ is the degree of the hub node. 
The degree of network synchronization can be characterized by the order 
parameter
\begin{equation} \label{eq:orderparm}
R=\langle|\frac{1}{N}\sum^{N}_{j=1}e^{i\theta_{j}}|\rangle_{T},
\end{equation}
where $j=1,\ldots,N$ is the node index, $N=4$ is the network size, $|\cdot|$ 
is the module function, and $\langle\cdot\rangle_{T}$ denotes the time average.

Setting $\omega=1$, we increase $\varepsilon$ systematically from $0.3$ to 
$0.8$, and calculate the dependence of $R$ on $\varepsilon$ by simulating 
Eq.~(\ref{eq:starnet}). In numerical simulations, the initial conditions of 
the oscillators are randomly chosen from the range $(0,2\pi]$ and the 
integration time step is $\Delta t=0.05$. Representative numerical results 
are shown in Fig.~\ref{fig:explosive}(b) (black solid squares). It can be 
seen that, at about $\varepsilon_{f}=0.55$, the value of $R$ changes 
abruptly from about $0.45$ to about $0.9$ - the feature of a discontinuous, 
first-order transition. The dynamical origin of this type of explosive 
synchronization lies in the interplay between the heterogeneity of the network 
structure and dynamics, which occurs naturally in networks where the node 
degree and the natural frequency are positively correlated~\cite{BAGLLSWZ:2016}.
Our goal is to use the reservoir machine trained by the time series from the 
desynchronization states to predict the critical coupling $\varepsilon_{f}$.

The data used in training consist of time series from five values of 
$\varepsilon$, all inside the desynchronization regime: 
$\varepsilon=0.4$, $0.425$, $0.45$, $0.475$ and $0.5$. For each $\varepsilon$
value, we collect time series measurements from all nodes: 
$x_j=\sin(\theta_j)$ and $y_j=\cos(\theta_j)$. The input state vector is 
$\mathbf{u}=[x_1,y_1,x_2,y_2,x_3,y_3,x_4,y_4]^T$, and we choose the data 
length to be $1.5\times 10^{4}$ that contains five segments of measurements,
one from each value of $\varepsilon$. These data, together with the parameter
function $\varepsilon(t)$, are fed into the machine for determining the output 
matrix $\mathcal{W}_{out}$. The parameters of the reservoir are 
$(D_r, p, \sigma, \rho,\alpha,\varepsilon_k,\varepsilon_{bias})=
(1000, 0.2, 1, 1.15,1,1,0)$, and the regression parameter 
is $\lambda=1\times 10^{-10}$. 

In the predicating phase, we increase the control parameter $\varepsilon$ 
systematically from $0.3$ to $0.8$ with the step $\Delta \varepsilon=0.01$, 
and calculate from the machine output the variation of the synchronization 
order parameter $R$, where the output vector 
$\mathbf{u}=[x_1,y_1,x_2,y_2,x_3,y_3,x_4,y_4]^T$ is transformed back to 
the original state variables $[\theta_1,\theta_2,\theta_3,\theta_4]^T$ 
through $\theta_i=\arctan (y_i/x_i)+\pi/2$ for $x_i\geq 0$ and 
$\theta_i=\arctan (y_i/x_i)+\pi$ for $x_i< 0$. The prediction results 
are shown in Fig.~\ref{fig:explosive}(b) (red open circles). It can be seen
that, at about $\varepsilon^{rc}_{f}=0.56$, the value of $R$ changes suddenly 
from about $0.38$ to about $0.8$ - the feature of a first-order transition. 

A distinct feature of explosive synchronization transitions is that when 
$\varepsilon$ is varied in the opposite direction, the variation of $R$ will 
follow a different path - the phenomenon of 
hysteresis~\cite{GGAM:2011,ZPSLK:2014,BAGLLSWZ:2016}. To demonstrate it,
we decrease $\varepsilon$ from $0.8$ to $0.3$. To numerically observe the
hysteresis in this context, we apply small random perturbations of amplitude 
$5\%$ to the natural frequency of the leaf nodes~\cite{ZPSLK:2014}. 
Figure~\ref{fig:explosive}(b) shows the results (blue solid up-triangles), 
where the backward and forward transition paths are identical for 
$\varepsilon>\varepsilon_f$, but diverge from each other for 
$\varepsilon \leq\varepsilon_f$. Particularly, at $\varepsilon_f$, we have 
$R\approx 0.9$ for the backward path, whereas $R\approx 0.38$ for the forward 
path. Along the backward path, as $\varepsilon$ decreases from $\varepsilon_f$,
the value of $R$ maintains at large values until the critical coupling 
$\varepsilon_{b}\approx 0.51$, where $R$ is suddenly decreased from about 
$0.8$ to about $0.4$. Since $\varepsilon_{b}<\varepsilon_{f}$, a hysteresis 
loop of width $\Delta \varepsilon \equiv \varepsilon_{f}-\varepsilon_{b}$ 
emerges in the parameter region 
$\varepsilon\in (\varepsilon_{b},\varepsilon_{f})$. The dynamical mechanism
for the hysteresis loop is the bistability of the synchronization manifold 
in this region, deemed as a necessary condition for generating a first-order 
phase transition~\cite{BAGLLSWZ:2016}. 

To predict the backward transition path, we use five values of the coupling
parameter in the strong synchronization regime to obtain the training data:
$\varepsilon=0.65$, $0.625$, $0.6$, $0.575$ and $0.55$. The input vector
is constructed in the same way as for predicting the forward transition, and
the hyperparameter values are $(D_r,p,\sigma,\rho)=(1\times 10^3,0.2,1,1.15)$ 
and the regression parameter is $\lambda=1\times 10^{-7}$. We decrease 
$\varepsilon$ systematically from $0.8$ to $0.3$. The dependence of $R$ on 
$\varepsilon$ predicted by the machine is shown in Fig.~\ref{fig:explosive}(b) 
(open green down-triangles). It can be seen that the machine predictions agree 
well with the true behavior of the backward transition where, at the transition
point $\varepsilon_b$, the value of $R$ decreases suddenly from about 0.8 to 
about 0.45.

\section{Discussion} \label{sec:discussion}

We have articulated and tested a model-free, machine learning scheme to predict
the synchronization transition in systems of coupled oscillators. The machine
is trained with time series collected from a small number of the coupling
(control) parameter, all in the desynchronization regime, as well as the value
of the control parameter itself through a specially designed input channel.
Prediction is achieved by feeding any desired parameter value into the input
parameter channel. A properly trained machine is able to not only reproduce,
statistically, the nature of the collective dynamics at the training parameter 
values, but also predict, quantitatively, how the collective dynamics change 
with respect to the variations in the control parameter. Examples demonstrating the predictive power of our machine learning scheme include complete 
synchronization in coupled identical chaotic oscillators and explosive 
synchronization in coupled nonidentical phase oscillators. For complete 
synchronization, both the critical coupling for synchronization and the 
variation in the degree of synchronization about the critical point can be well
predicted. For explosive synchronization, our scheme not only predicts the 
forward and backward critical couplings, but also reproduces the hysteresis 
loop associated with a first-order transition. Due to the importance of 
synchronization to the functionality and operation in many natural and 
man-made systems, our machine learning method may find broad applications.

Reservoir computing based prediction of chaotic systems is an extremely 
active field of research at the boundary between nonlinear dynamics and 
machine learning~\cite{LBE:2015,HSRFG:2015,LBMUCJ:2017,PLHGO:2017,LPHGBO:2017,DBN:book,LHO:2018,PWFCHGO:2018,PHGLO:2018,Carroll:2018,NS:2018,ZP:2018,WYGZS:2019,GPG:2019,JL:2019,VPHSGOK:2019,FJZWL:2020,ZJQL:2020,GYL:2021}.
The main contribution of the present work is the development of a reservoir 
computing scheme to predict or anticipate synchronization in systems of 
coupled nonlinear oscillators. Our work focuses on predicting the collective 
dynamics instead of the state evolution, based on the general idea to view the 
predictive power of reservoir computing as a kind of ability to replicate the 
dynamical ``climate'' of the target system~\cite{PLHGO:2017}, which is gained 
through training with time series data. Inspired by the recent works on 
conducting training at multiple parameter values to predict 
bifurcations~\cite{CA:2019,KFGL:2021}, our work adopts a similar method by 
training the machine at a small number of control parameter values in the 
desynchronization regime to instill into the machine the ability to sense the 
change in the ``synchronization climate'' with the control parameter. With the 
desired (arbitrary) value of the control parameter fed into the input parameter
channel, a well trained reservoir computing machine is then able to accurately 
predict the critical transition between desynchronization and synchronization, 
regardless of the nature of the underlying transition, e.g., second-order or 
first-order.

The type of collective dynamics tested in the present work is complete
synchronization between a pair of coupled chaotic oscillators for which the
transition is of the nature of second order, and the first order, explosive 
synchronization in a small network. To extend our work to other types of 
collective dynamics in large complex systems, such as partial (cluster) 
synchronization, chimera-like states and spiral waves, is worth pursuing. A 
difficulty with large systems is the requirement to use large reservoir 
networks so that the complexity of the machine can ``overpower'' that of the 
target system. Quantitatively, how the size of the reservoir network should 
be enlarged to accommodate an increase in the size of the target system as 
characterized by, e.g., a scaling law, remains unknown at the present. With 
the use of large reservoir networks come the issues of data requirement and 
computation overload, as to train a large reservoir machine not only requires 
massive data but also imposes a serious demand for the computational resource. 
One approach to deal with this difficulty is the parallel reservoir computing 
scheme~\cite{PHGLO:2018,ZP:2018}. However, a recent work revealed that the 
parallel scheme may fail to sense and predict the phase coherence among a 
pair of coupled, nonidentical chaotic oscillators~\cite{ZJQL:2020}. It 
remains a worthy issue to study if the parallel scheme can be exploited to 
predict the collective dynamics among a large number of coupled oscillators.

\begin{acknowledgments}

This work was supported by the National Natural Science Foundation of China 
under the Grant No.~11875182. The work at Arizona State University was 
supported by the Pentagon Vannevar Bush Faculty Fellowship program sponsored 
by the Basic Research Office of the Assistant Secretary of Defense for 
Research and Engineering and funded by the Office of Naval Research through 
Grant No.~N00014-16-1-2828, and by the Army Research Office through 
Grant No.~W911NF2120055.

\end{acknowledgments}

\appendix

\section{Robustness of reservoir computing scheme for predicting synchronization transition} \label{appendix_a}

\subsection{Effect of nonuniform training data on predication performance}

\begin{figure}[ht!]
\centering
\includegraphics[width=\linewidth]{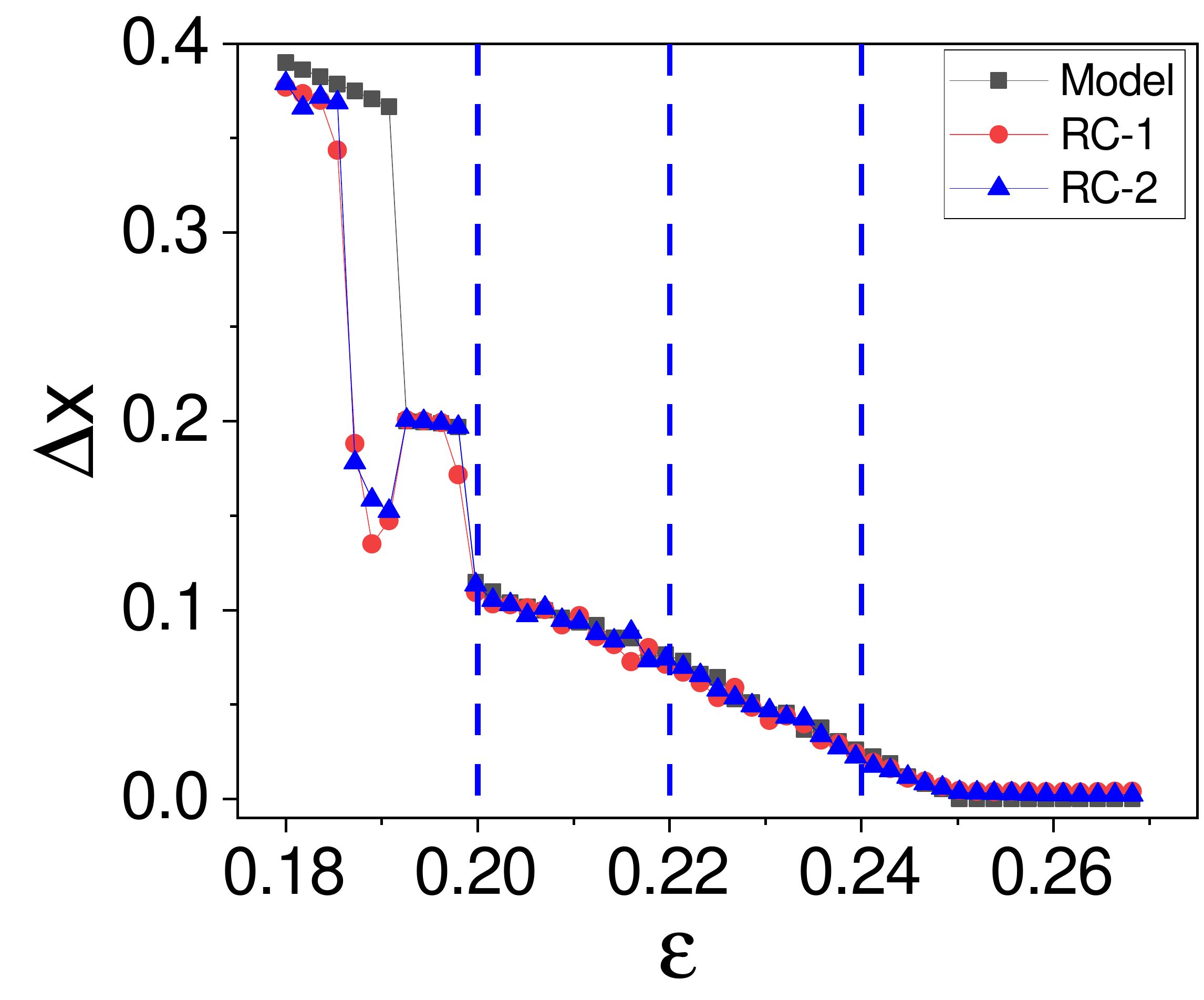}
\caption{ Effect of nonuniform segment length on predication performance.
Shown is the variation of the synchronization error $\Delta x$ versus the
coupling parameter $\varepsilon$ for the system of coupled chaotic logistic
maps in the main text. The parameter values are the same as those in Fig.~2(a)
in the main text, but with nonuniform length of the time series for different
values of the training parameter. Red discs: $T=1000$, $2000$ and $3000$ for
$\varepsilon=0.2$, $0.22$ and $0.24$, respectively; blue triangles: $T=3000$,
$2000$ and $1000$ for $\varepsilon=0.2$, $0.22$ and $0.24$, respectively.
The black squares denote the corresponding results from the target system.
The vertical dashed lines denote the values of the control parameter for
training.}
\label{fig:different_Ti}
\end{figure}

In the main text, the lengths of the time series for training at different
values of the control parameter are identical. For example, for the system
of a pair of chaotic logistic maps in Sec.~IIIA, the length of the time
series is $T=2\times 10^3$. If time series of nonuniform length are used
for training, the predication performance will be not affected. To demonstrate
this, we calculate the machine predicted synchronization error versus the
coupling parameter using training time series of length $T=1000$, $2000$ and
$3000$ for $\varepsilon=0.2$, $0.22$ and $0.24$, respectively, as shown
in Fig.~\ref{fig:different_Ti} (red discs). Comparing with the results of
from direct model simulations (black squares), we see that the reservoir
machine still predicts well the transition about the critical point. Similar
results are also obtained when the training data have the length $T=3000$ for
$\varepsilon=0.20$, $2000$ for $\varepsilon=0.22$, and $1000$ for
$\varepsilon=0.24$, as shown by the blue triangles in
Fig.~\ref{fig:different_Ti}. The observation is that, insofar as the training
data set is sufficient (e.g., $>1000$), both the critical point and the
associated transition behavior can be well predicted.

\subsection{Effect of order of training at different control parameter values on predication performance}

\begin{figure}[ht!]
\centering
\includegraphics[width=\linewidth]{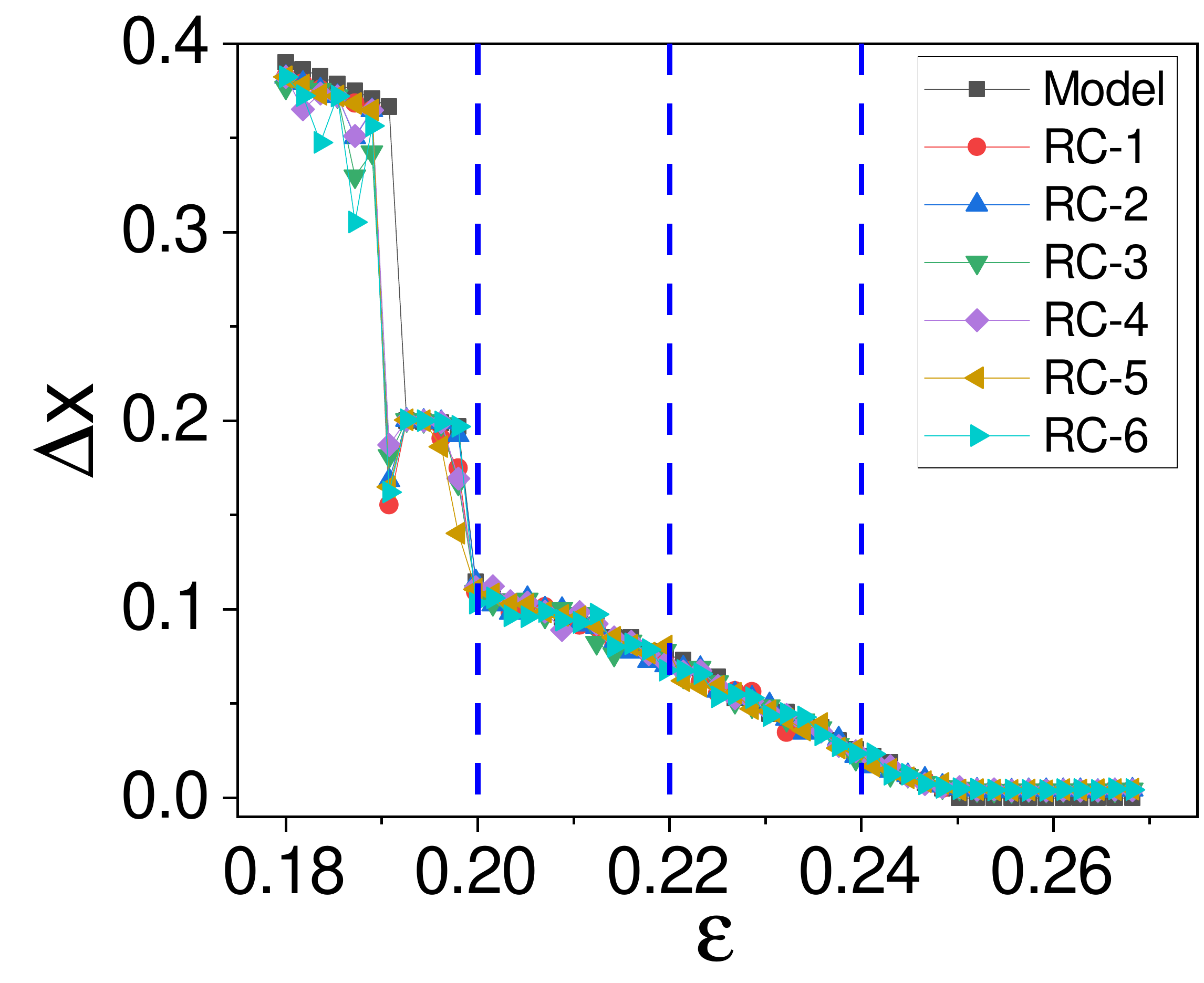}
\caption{Effect of the order of training at different control parameter values
on predication performance. The model and parameters are identical to those
in Fig.~2(a) in the main text, while the order of training at the three
values of the control parameter ($\varepsilon=0.2$, $0.22$ and $0.24$, as
indicated by the vertical dashed lines) is altered. Shown is the
synchronization error $\Delta x$ versus $\varepsilon$. A change in the
training order has little effect on the performance.}
\label{fig:permuntation_Ti}
\end{figure}

In our machine learning scheme, the training data are generated by combining
the time series acquired from different values of the control parameter. We
find that order of training at these parameter values does not affect the
the predication performance. To demonstrate this, we use the results in
Fig.~2(a) in the main text as a reference and obtain prediction results for
altered training orders. Since data are obtained from three values of the
control parameter $\varepsilon$ (0.2, 0.22, and 0.24), there are six distinct
sequences of training order. Figure~\ref{fig:permuntation_Ti} shows the
machine predicted synchronization error $\Delta x$ versus $\varepsilon$ for
all six cases, together with the true results obtained from direct simulation
of the target system. It can be seen that different orders of training lead
to essentially the same result. A heuristic reason is that, in obtaining the
output matrix $\mathbf{W}_{out}$, the regression operation is conducted for
the entire, combined time series. For a linear regression, the orders by which
the time series are combined have little effect on the result.

\subsection{Impact of training control parameter values on predication performance}

\begin{figure}[tbp]
\centering
\includegraphics[width=\linewidth]{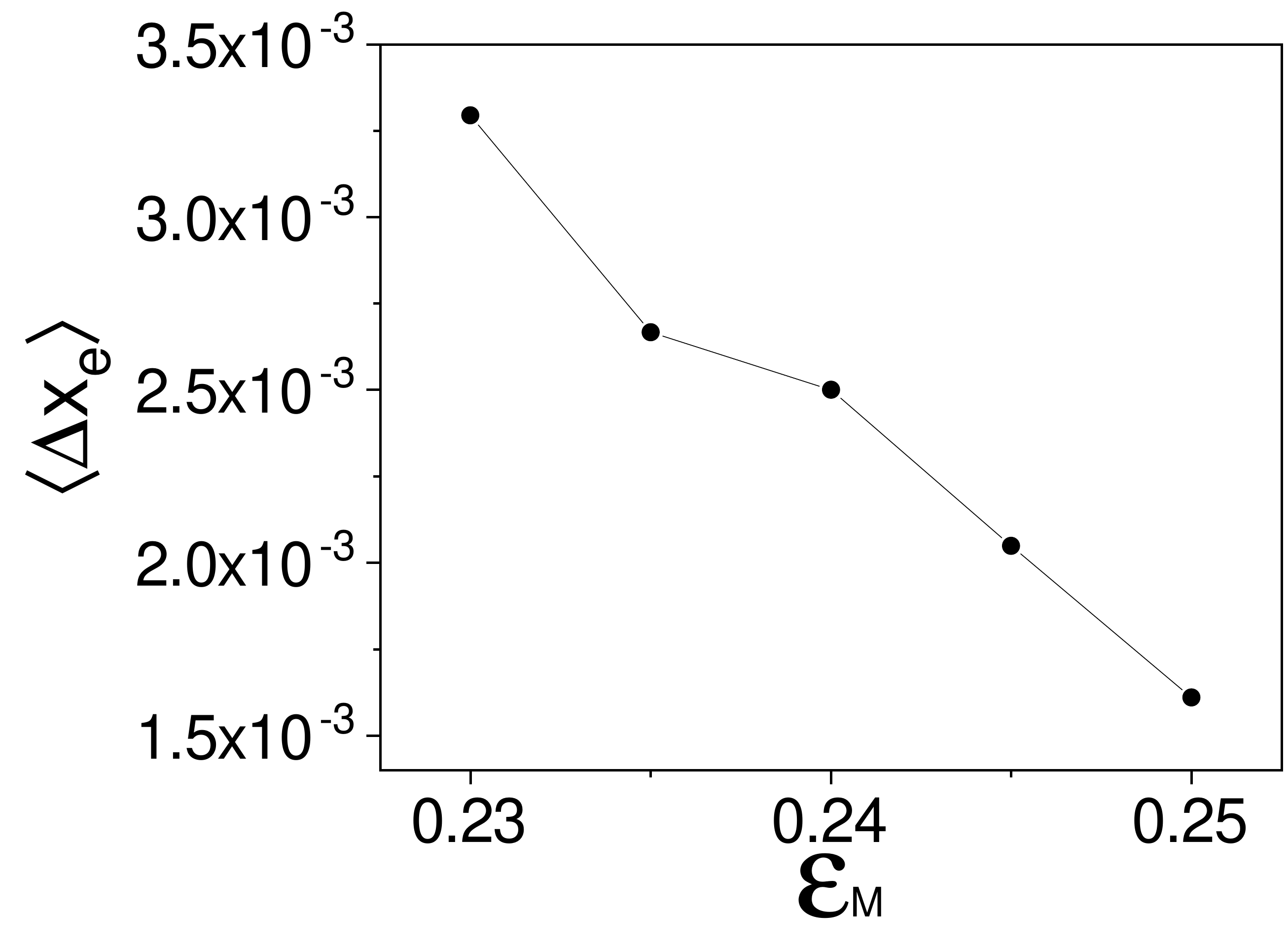}
\caption{ Impact of training control parameter values on predication
performance. With the model for Fig.~2(a) in the main text, the values of
the control parameter for training ($\varepsilon=0.2$, $0.22$ and $0.24$)
are shifted by a small amount toward the critical point $\varepsilon_1=0.25$.
Shown is the predication error $\langle\Delta x_{e}\rangle$ versus the
parameter value for training, denoted as $\varepsilon_M$. As the training
parameter values move closer to the critical point (decreasing $\varepsilon_M$), the predication
performance is continuously improved.}
\label{fig:shift_training_state}
\end{figure}

For the given number of training parameter values, the closer they are to
the transition point, the better the predication performance will be. To
demonstrate this, we consider the system of coupled chaotic Logistic maps
in the main text and test the predication performance by changing the values 
of $\varepsilon$ to $\varepsilon_M-0.04$, $\varepsilon_M-0.02$ and 
$\varepsilon_M$, where $\varepsilon_M$ is below the synchronization transition 
point so that all three parameter values are still in the desynchronization 
regime. We define the following prediction error about the transition as
\begin{displaymath}
\langle\Delta x_{e}\rangle=\langle |\Delta \hat{x}(\varepsilon)-\Delta x(\varepsilon)|\rangle,
\end{displaymath}
where $\Delta \hat{x}(\varepsilon)$ and $\Delta x(\varepsilon)$ are the
synchronization errors obtained from model simulation and machine outputs,
and $\langle\ldots\rangle$ denotes that the results are averaged over the
range $0.245 \le \varepsilon \le 0.255$. Figure~\ref{fig:shift_training_state}
shows $\langle\Delta x_{e}\rangle$ versus $\varepsilon_M$, the largest training
parameter value chosen in the desynchronization regime. As $\varepsilon_M$ 
approaches the critical point, the error $\langle\Delta x_{e}\rangle$ 
decreases continuously, indicating that choosing the training parameter values
closer to the critical point can improve the prediction.

\subsection{Impact of the number of training parameter values on predication performance}

\begin{figure}[tbp]
\centering
\includegraphics[width=\linewidth]{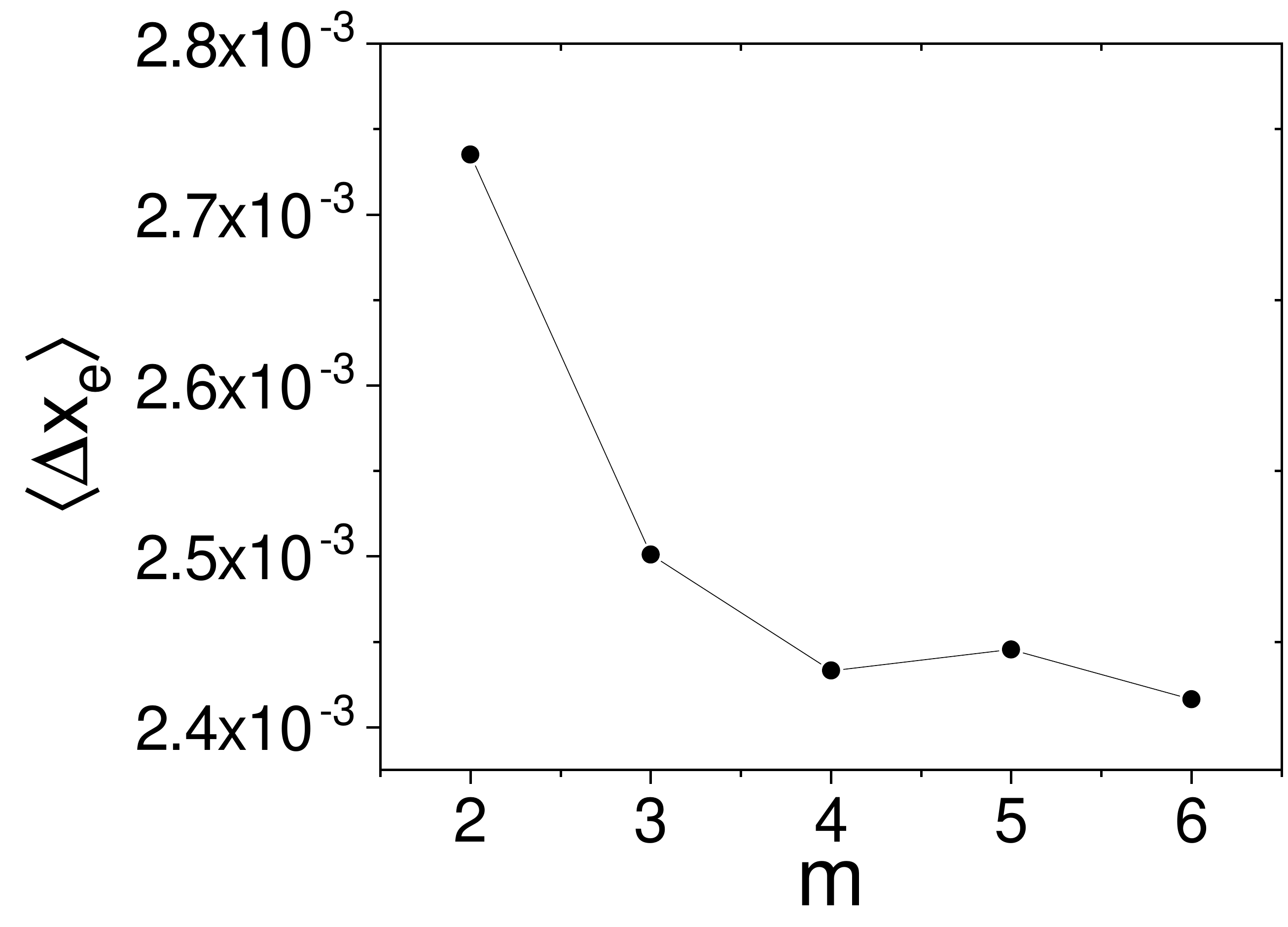}
\caption{Impact of the number of training parameter values on predication
performance. The model and parameter values are identical to those in
Fig.~2(a) in the main text, except that the number $m$ of control parameter
values chosen to obtain the time series data for training now varies.
Shown is the predication error $\langle\Delta x_{e}\rangle$ versus $m$.
An increase in $m$ above three improves the predication performance but only
slightly, indicating that the reservoir computing machine is already able
to capture the dynamical ``climate'' of the target system with training
conducted at three values of the control parameter.}
\label{fig:training_states_n}
\end{figure}

Increasing the number $m$ of control parameter values in the desynchronization
regime for training can in general improve the prediction performance, but
only incrementally, insofar as training is conducted based on time series
data collected from at least two values of the control parameter. To
demonstrate this, we take the model in Fig.~2(a) in the main text and
investigate how the value of $m$ impacts the predicted synchronization error,
For simplicity, we choose the $m$ training parameter values from the interval
$[0.20,0.24]$ (including the two ending points). 
Figure~\ref{fig:training_states_n} shows $\langle\Delta x_{e}\rangle$ versus 
$m$. As the value of $m$ increases from $2$ to $4$, the predication error is 
reduced but the trend slows down for $m > 4$, indicating that it is generally 
not necessary to conduct the training for more than three or four values of the control parameter.

\section{Alternative types of coupling functions} \label{appendix_b}

\begin{figure}[tbp]
\centering
\includegraphics[width=\linewidth]{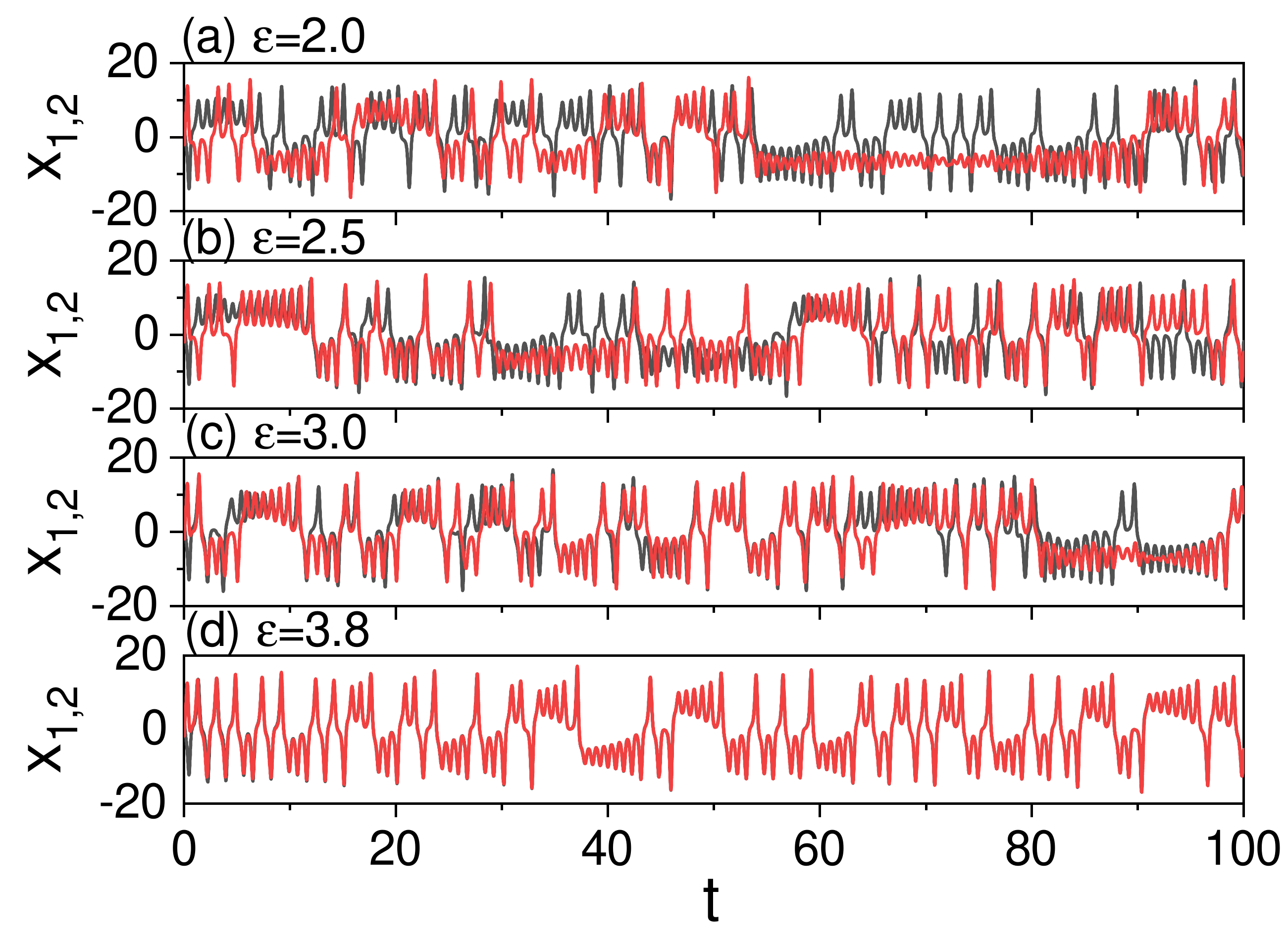}
\caption{ Predicting collective dynamical behaviors arising from the system
of two coupled Lorenz chaotic oscillators under the $x$ coupling scheme.
Shown is the time evolution of the reservoir outputs $x_1$ (black curve)
and $x_2$ (red curve) for different values of the control parameter:
(a) $\varepsilon=2.0$, (b) $\varepsilon=2.5$, (c) $\varepsilon=3.0$,
and (d) $\varepsilon=3.8$. There is no synchronization in the first three
cases, but there is for the last case, in agreement with the result of the
analysis based on the master stability function that the oscillators are
completely synchronized for $\varepsilon>\varepsilon_c=3.7$.}
\label{fig:xx_coupling}
\end{figure}

The coupling function can play a role in the synchronization dynamics:
changing the function can alter the synchronization type and the critical
point. In the main text, we treat the relatively simple setting of diagonal
coupling function, i.e., all the dynamical variables of the oscillators are
diffusively coupled. To test if the machine learning scheme remains effective
for alternative types of coupling functions, we take the model of coupled
chaotic Lorenz oscillators studied in the main text [Fig.~\ref{fig:Lorenz}(a)] 
and evaluate the prediction performance with different coupling functions.

\begin{figure}[tbp]
\centering
\includegraphics[width=\linewidth]{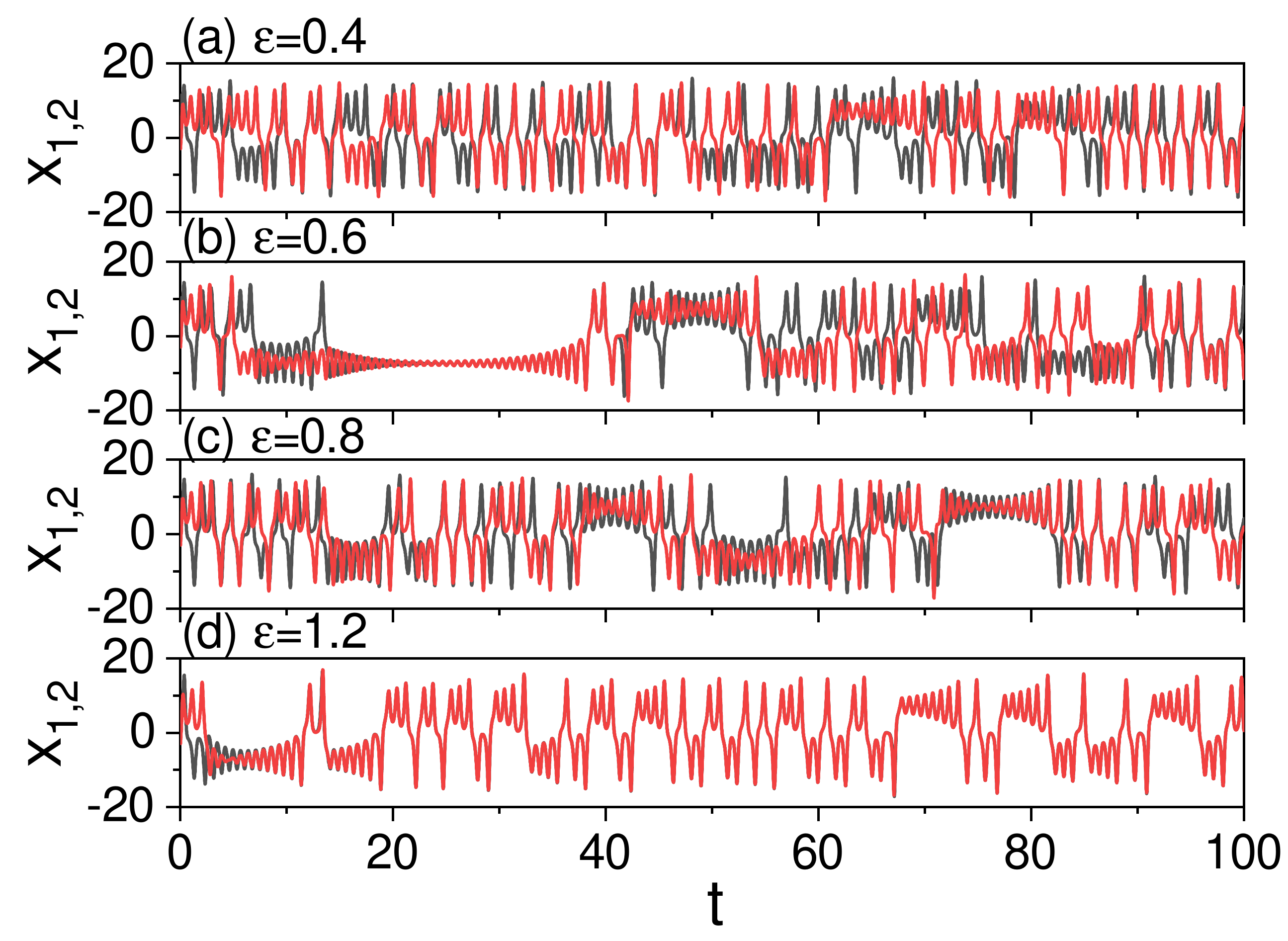}
\caption{ Predicting collective dynamical behaviors arising from the system
of two coupled Lorenz chaotic oscillators under the $y$ coupling scheme.
Legends are the same as those in Fig.~\ref{fig:xx_coupling}.
For (a) $\varepsilon=0.4$, (b) $\varepsilon=0.6$, (c) $\varepsilon=0.8$,
there is no synchronization, but complete synchronization occurs for
(d) $\varepsilon=1.2$, in agreement with the prediction of the theoretical
analysis that the oscillators should be completely synchronized for
$\varepsilon>\varepsilon_c=3.7$.}
\label{fig:yy_coupling}
\end{figure}

When the oscillators are coupled through the variable $x$ of the Lorenz
oscillator, the system equations are
\begin{align} \label{lorenz-x}
&\dot{x}_{1,2}=\alpha (y_{1,2}-x_{1,2})+\varepsilon(x_{2,1}-x_{1,2}) \notag, \\
&\dot{y}_{1,2}=x_{1,2}(\beta-z_{1,2})-y_{1,2}, \\
&\dot{z}_{1,2}=x_{1,2}y_{1,2}-\gamma z_{1,2}\notag.
\end{align}
We use the same parameter values as in the main text. An analysis of the
master stability function reveals that, for this alternative coupling scheme,
the two oscillators will be completely synchronized for
$\varepsilon>\varepsilon_c\approx 3.7$. To generate the training data, we
collect time series from three values of the control parameter:
$\varepsilon=2.0$, $2.5$ and $3.0$. The parameters of the reservoir are
$(D_r, p, \sigma, \rho, \alpha, \varepsilon_k, \varepsilon)=(2\times 10^3, 0.2, 0.05, 1\times 10^{-3}, 1,1,0)$ and the
regression parameter is $\lambda = 10^{-7}$. Fig.~\ref{fig:xx_coupling} shows
the prediction result, where the machine output variables $x_1$ and $x_2$
are not synchronized for $\varepsilon=2.0$, 2.5, and 3.0
[Figs.~\ref{fig:xx_coupling}(a-c), respectively], but are synchronized for
$\varepsilon=3.8$ [Fig.~\ref{fig:xx_coupling}(d)], in complete agreement
with the theoretical analysis.

We next consider the scheme by which the $y$ variables of the oscillators
are coupled. The system equations are
\begin{align}\label{lorenz-y}
&\dot{x}_{1,2}=\alpha (y_{1,2}-x_{1,2}) \notag, \\
&\dot{y}_{1,2}=x_{1,2}(\beta-z_{1,2})-y_{1,2}+\varepsilon(y_{2,1}-y_{1,2}), \\
&\dot{z}_{1,2}=x_{1,2}y_{1,2}-\gamma z_{1,2}\notag,
\end{align}
where the two oscillators are completely synchronized for
$\varepsilon>\varepsilon_c\approx 1.1$, as predicated by the master stability
function. We train the reservoir machine based on data from $\varepsilon=0.4$,
$0.6$ and 0.8. The parameter values of the reservoir are the same as for
Figs.~\ref{fig:xx_coupling}(a-d), but the regression parameter is
$\lambda = 10^{-10}$. Figure~\ref{fig:yy_coupling} shows the predicted
synchronization behavior between the machine outputs $x_1$ and $x_2$ for
different values of the control parameter. In complete agreement with the
result of the theoretical analysis, there is no synchronization for
$\varepsilon=0.4$, 0.6, and 0.8 [Figs.~\ref{fig:yy_coupling}(a-c),
respectively], but the machine output variables are synchronized
for $\varepsilon=1.2>\varepsilon_c$, as shown in Fig.~\ref{fig:yy_coupling}(d).

\begin{figure}[tbp]
\centering
\includegraphics[width=\linewidth]{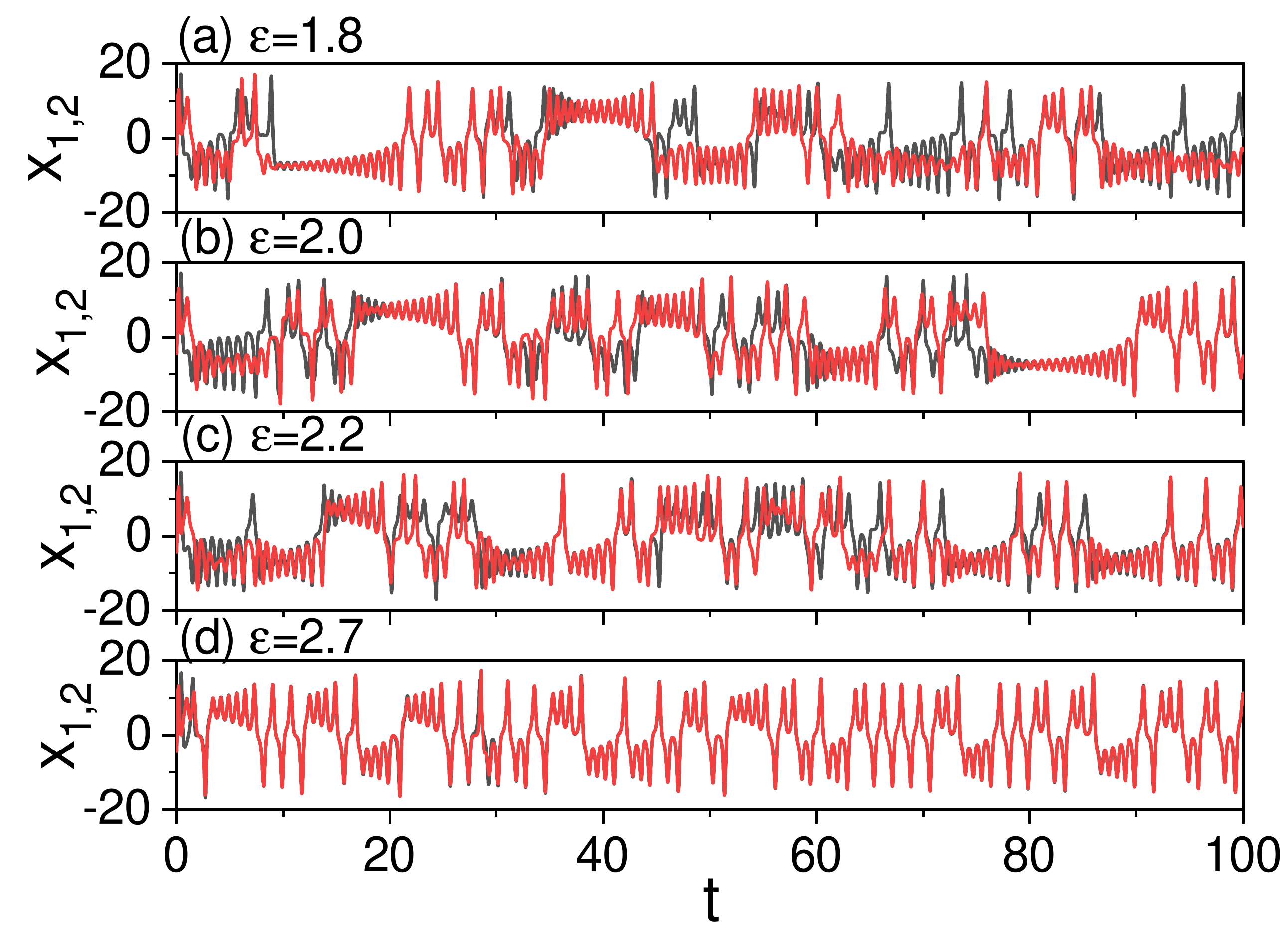}
\caption{Predicting collective dynamical behaviors arising from the system
of two coupled Lorenz chaotic oscillators under the cross coupling scheme.
Legends are the same as those in Fig.~\ref{fig:xx_coupling}.
For (a) $\varepsilon=1.8$, (b) $\varepsilon=2.0$, (c) $\varepsilon=2.28$,
there is no synchronization, but complete synchronization occurs for
(d) $\varepsilon=2.7$, in agreement with the prediction of the theoretical
analysis that the oscillators should be completely synchronized for
$\varepsilon>\varepsilon_c=2.6$.}
\label{fig:xy_coupling}
\end{figure}

We finally consider the cross coupling scheme, i.e., variable $x$ of one
oscillator is coupled to variable $y$ of the other. The system is described by
\begin{align}\label{lorenz-cross}
&\dot{x}_{1,2}=\alpha (y_{1,2}-x_{1,2}) \notag, \\
&\dot{y}_{1,2}=x_{1,2}(\beta-z_{1,2})-y_{1,2}+\varepsilon(x_{2,1}-x_{1,2}), \\
&\dot{z}_{1,2}=x_{1,2}y_{1,2}-\gamma z_{1,2}\notag.
\end{align}
For this coupling scheme, according to the theoretical analysis, the two
oscillators are completely synchronized for
$\varepsilon>\varepsilon_c\approx 2.6$. Training data are generated for
$\varepsilon=1.8$, $2.0$ and $2.2$, all in the desynchronization regime.
The values of the hyperparameters of the reservoir and of the regression
parameter are the same as those in Fig.~\ref{fig:yy_coupling}.
Figure~\ref{fig:xy_coupling} shows the predicted synchronization behavior
between the reservoir outputs $x_1$ and $x_2$ for different values of
$\varepsilon$. There is no synchronization for $\varepsilon=1.8$, 2.0, and
2.2, as shown in Fig.~\ref{fig:xy_coupling}(a-c), respectively, but for
$\varepsilon=2.7>\varepsilon_c$, the machine predicts synchronization, as
shown in Fig.~\ref{fig:xy_coupling}(d). These machine predicted behaviors
are in full agreement with the result of the theoretical analysis.


\begin{thebibliography}{56}%
\makeatletter
\providecommand \@ifxundefined [1]{%
 \@ifx{#1\undefined}
}%
\providecommand \@ifnum [1]{%
 \ifnum #1\expandafter \@firstoftwo
 \else \expandafter \@secondoftwo
 \fi
}%
\providecommand \@ifx [1]{%
 \ifx #1\expandafter \@firstoftwo
 \else \expandafter \@secondoftwo
 \fi
}%
\providecommand \natexlab [1]{#1}%
\providecommand \enquote  [1]{``#1''}%
\providecommand \bibnamefont  [1]{#1}%
\providecommand \bibfnamefont [1]{#1}%
\providecommand \citenamefont [1]{#1}%
\providecommand \href@noop [0]{\@secondoftwo}%
\providecommand \href [0]{\begingroup \@sanitize@url \@href}%
\providecommand \@href[1]{\@@startlink{#1}\@@href}%
\providecommand \@@href[1]{\endgroup#1\@@endlink}%
\providecommand \@sanitize@url [0]{\catcode `\\12\catcode `\$12\catcode
  `\&12\catcode `\#12\catcode `\^12\catcode `\_12\catcode `\%12\relax}%
\providecommand \@@startlink[1]{}%
\providecommand \@@endlink[0]{}%
\providecommand \url  [0]{\begingroup\@sanitize@url \@url }%
\providecommand \@url [1]{\endgroup\@href {#1}{\urlprefix }}%
\providecommand \urlprefix  [0]{URL }%
\providecommand \Eprint [0]{\href }%
\providecommand \doibase [0]{http://dx.doi.org/}%
\providecommand \selectlanguage [0]{\@gobble}%
\providecommand \bibinfo  [0]{\@secondoftwo}%
\providecommand \bibfield  [0]{\@secondoftwo}%
\providecommand \translation [1]{[#1]}%
\providecommand \BibitemOpen [0]{}%
\providecommand \bibitemStop [0]{}%
\providecommand \bibitemNoStop [0]{.\EOS\space}%
\providecommand \EOS [0]{\spacefactor3000\relax}%
\providecommand \BibitemShut  [1]{\csname bibitem#1\endcsname}%
\let\auto@bib@innerbib\@empty
\bibitem [{\citenamefont {Kuramoto}(1984)}]{Kuramoto:book}%
  \BibitemOpen
  \bibfield  {author} {\bibinfo {author} {\bibfnamefont {Y.}~\bibnamefont
  {Kuramoto}},\ }\href@noop {} {\emph {\bibinfo {title} {Chemical Oscillations,
  Waves, and Turbulence}}}\ (\bibinfo  {publisher} {Springer, Berlin},\
  \bibinfo {year} {1984})\BibitemShut {NoStop}%
\bibitem [{\citenamefont {Pikovsky}\ \emph {et~al.}(2003)\citenamefont
  {Pikovsky}, \citenamefont {Rosenblum},\ and\ \citenamefont
  {Kurths}}]{PRK:book}%
  \BibitemOpen
  \bibfield  {author} {\bibinfo {author} {\bibfnamefont {A.}~\bibnamefont
  {Pikovsky}}, \bibinfo {author} {\bibfnamefont {M.}~\bibnamefont {Rosenblum}},
  \ and\ \bibinfo {author} {\bibfnamefont {J.}~\bibnamefont {Kurths}},\
  }\href@noop {} {\emph {\bibinfo {title} {Synchronization: A Universal Concept
  in Nonlinear Science}}}\ (\bibinfo  {publisher} {Cambridge University Press,
  Cambridge},\ \bibinfo {year} {2003})\BibitemShut {NoStop}%
\bibitem [{\citenamefont {Strogatz}(2003)}]{Strogatz:book}%
  \BibitemOpen
  \bibfield  {author} {\bibinfo {author} {\bibfnamefont {S.~H.}\ \bibnamefont
  {Strogatz}},\ }\href@noop {} {\emph {\bibinfo {title} {Sync: The Emerging
  Science of Spontaneous Order}}}\ (\bibinfo  {publisher} {Hyperion, New
  York},\ \bibinfo {year} {2003})\BibitemShut {NoStop}%
\bibitem [{\citenamefont {Pecora}\ and\ \citenamefont
  {Carroll}(1990)}]{PC:1990}%
  \BibitemOpen
  \bibfield  {author} {\bibinfo {author} {\bibfnamefont {L.~M.}\ \bibnamefont
  {Pecora}}\ and\ \bibinfo {author} {\bibfnamefont {T.~L.}\ \bibnamefont
  {Carroll}},\ }\bibfield  {title} {\enquote {\bibinfo {title} {Synchronization
  in chaotic systems},}\ }\href {\doibase 10.1103/PhysRevLett.64.821}
  {\bibfield  {journal} {\bibinfo  {journal} {Phys. Rev. Lett.}\ }\textbf
  {\bibinfo {volume} {64}},\ \bibinfo {pages} {821} (\bibinfo {year}
  {1990})}\BibitemShut {NoStop}%
\bibitem [{\citenamefont {Rosenblum}\ \emph {et~al.}(1996)\citenamefont
  {Rosenblum}, \citenamefont {Pikovsky},\ and\ \citenamefont
  {Kurths}}]{RPK:1996}%
  \BibitemOpen
  \bibfield  {author} {\bibinfo {author} {\bibfnamefont {M.~G.}\ \bibnamefont
  {Rosenblum}}, \bibinfo {author} {\bibfnamefont {A.~S.}\ \bibnamefont
  {Pikovsky}}, \ and\ \bibinfo {author} {\bibfnamefont {J.}~\bibnamefont
  {Kurths}},\ }\bibfield  {title} {\enquote {\bibinfo {title} {Phase
  synchronization of chaotic oscillators},}\ }\href {\doibase
  10.1103/PhysRevLett.76.1804} {\bibfield  {journal} {\bibinfo  {journal}
  {Phys. Rev. Lett.}\ }\textbf {\bibinfo {volume} {76}},\ \bibinfo {pages}
  {1804} (\bibinfo {year} {1996})}\BibitemShut {NoStop}%
\bibitem [{\citenamefont {Kocarev}\ and\ \citenamefont
  {Parlitz}(1996)}]{KP:1996}%
  \BibitemOpen
  \bibfield  {author} {\bibinfo {author} {\bibfnamefont {L.}~\bibnamefont
  {Kocarev}}\ and\ \bibinfo {author} {\bibfnamefont {U.}~\bibnamefont
  {Parlitz}},\ }\bibfield  {title} {\enquote {\bibinfo {title} {Generalized
  synchronization, predictability, and equivalence of unidirectionally coupled
  dynamical systems},}\ }\href {\doibase 10.1103/PhysRevLett.76.1816}
  {\bibfield  {journal} {\bibinfo  {journal} {Phys. Rev. Lett.}\ }\textbf
  {\bibinfo {volume} {76}},\ \bibinfo {pages} {1816} (\bibinfo {year}
  {1996})}\BibitemShut {NoStop}%
\bibitem [{\citenamefont {Kandel}\ \emph {et~al.}(1991)\citenamefont {Kandel},
  \citenamefont {Schwartz},\ and\ \citenamefont {Jessell}}]{KSJ:book}%
  \BibitemOpen
  \bibfield  {author} {\bibinfo {author} {\bibfnamefont {E.~R.}\ \bibnamefont
  {Kandel}}, \bibinfo {author} {\bibfnamefont {J.~H.}\ \bibnamefont
  {Schwartz}}, \ and\ \bibinfo {author} {\bibfnamefont {T.~M.}\ \bibnamefont
  {Jessell}},\ }\href@noop {} {\emph {\bibinfo {title} {{Principle of Neural
  Science}}}},\ \bibinfo {edition} {3rd}\ ed.\ (\bibinfo  {publisher} {Appleton
  and Lange},\ \bibinfo {address} {Norwalk CT},\ \bibinfo {year}
  {1991})\BibitemShut {NoStop}%
\bibitem [{\citenamefont {Lai}\ \emph {et~al.}(2006)\citenamefont {Lai},
  \citenamefont {Frei},\ and\ \citenamefont {Osorio}}]{LFO:2006}%
  \BibitemOpen
  \bibfield  {author} {\bibinfo {author} {\bibfnamefont {Y.-C.}\ \bibnamefont
  {Lai}}, \bibinfo {author} {\bibfnamefont {M.~G.}\ \bibnamefont {Frei}}, \
  and\ \bibinfo {author} {\bibfnamefont {I.}~\bibnamefont {Osorio}},\
  }\bibfield  {title} {\enquote {\bibinfo {title} {Detecting and characterizing
  phase synchronization in nonstationary dynamical systems},}\ }\href {\doibase
  10.1103/PhysRevE.73.026214} {\bibfield  {journal} {\bibinfo  {journal} {Phys.
  Rev. E}\ }\textbf {\bibinfo {volume} {73}},\ \bibinfo {pages} {026214}
  (\bibinfo {year} {2006})}\BibitemShut {NoStop}%
\bibitem [{\citenamefont {Lai}\ \emph {et~al.}(2007)\citenamefont {Lai},
  \citenamefont {Frei}, \citenamefont {Osorio},\ and\ \citenamefont
  {Huang}}]{LFOH:2007}%
  \BibitemOpen
  \bibfield  {author} {\bibinfo {author} {\bibfnamefont {Y.-C.}\ \bibnamefont
  {Lai}}, \bibinfo {author} {\bibfnamefont {M.~G.}\ \bibnamefont {Frei}},
  \bibinfo {author} {\bibfnamefont {I.}~\bibnamefont {Osorio}}, \ and\ \bibinfo
  {author} {\bibfnamefont {L.}~\bibnamefont {Huang}},\ }\bibfield  {title}
  {\enquote {\bibinfo {title} {Characterization of synchrony with applications
  to epileptic brain signals},}\ }\href@noop {} {\bibfield  {journal} {\bibinfo
   {journal} {Phys. Rev. Lett.}\ }\textbf {\bibinfo {volume} {98}},\ \bibinfo
  {pages} {108102} (\bibinfo {year} {2007})}\BibitemShut {NoStop}%
\bibitem [{\citenamefont {Osorio}\ and\ \citenamefont {Lai}(2011)}]{OL:2011}%
  \BibitemOpen
  \bibfield  {author} {\bibinfo {author} {\bibfnamefont {I.}~\bibnamefont
  {Osorio}}\ and\ \bibinfo {author} {\bibfnamefont {Y.-C.}\ \bibnamefont
  {Lai}},\ }\bibfield  {title} {\enquote {\bibinfo {title} {A
  phase-synchronization and random-matrix based approach to multichannel
  time-series analysis with application to epilepsy},}\ }\href@noop {}
  {\bibfield  {journal} {\bibinfo  {journal} {Chaos}\ }\textbf {\bibinfo
  {volume} {21}},\ \bibinfo {pages} {033108} (\bibinfo {year}
  {2011})}\BibitemShut {NoStop}%
\bibitem [{\citenamefont {Jaeger}(2001)}]{Jaeger:2001}%
  \BibitemOpen
  \bibfield  {author} {\bibinfo {author} {\bibfnamefont {H.}~\bibnamefont
  {Jaeger}},\ }\bibfield  {title} {\enquote {\bibinfo {title} {The ``echo
  state" approach to analysing and training recurrent neural networks-with an
  erratum note},}\ }\href@noop {} {\bibfield  {journal} {\bibinfo  {journal}
  {Bonn, Germany: German National Research Center for Information Technology
  GMD Technical Report}\ }\textbf {\bibinfo {volume} {148}},\ \bibinfo {pages}
  {13} (\bibinfo {year} {2001})}\BibitemShut {NoStop}%
\bibitem [{\citenamefont {Jaeger}\ and\ \citenamefont {Haas}(2004)}]{JH:2004}%
  \BibitemOpen
  \bibfield  {author} {\bibinfo {author} {\bibfnamefont {H.}~\bibnamefont
  {Jaeger}}\ and\ \bibinfo {author} {\bibfnamefont {H.}~\bibnamefont {Haas}},\
  }\bibfield  {title} {\enquote {\bibinfo {title} {Harnessing nonlinearity:
  Predicting chaotic systems and saving energy in wireless communication},}\
  }\href@noop {} {\bibfield  {journal} {\bibinfo  {journal} {Science}\ }\textbf
  {\bibinfo {volume} {304}},\ \bibinfo {pages} {78} (\bibinfo {year}
  {2004})}\BibitemShut {NoStop}%
\bibitem [{\citenamefont {Pecora}\ and\ \citenamefont
  {Carroll}(1998)}]{PC:1998}%
  \BibitemOpen
  \bibfield  {author} {\bibinfo {author} {\bibfnamefont {L.~M.}\ \bibnamefont
  {Pecora}}\ and\ \bibinfo {author} {\bibfnamefont {T.~L.}\ \bibnamefont
  {Carroll}},\ }\bibfield  {title} {\enquote {\bibinfo {title} {Master
  stability functions for synchronized coupled systems},}\ }\href {\doibase
  10.1103/PhysRevLett.80.2109} {\bibfield  {journal} {\bibinfo  {journal}
  {Phys. Rev. Lett.}\ }\textbf {\bibinfo {volume} {80}},\ \bibinfo {pages}
  {2109} (\bibinfo {year} {1998})}\BibitemShut {NoStop}%
\bibitem [{\citenamefont {Huang}\ \emph {et~al.}(2009)\citenamefont {Huang},
  \citenamefont {Chen}, \citenamefont {Lai},\ and\ \citenamefont
  {Pecora}}]{HCLP:2009}%
  \BibitemOpen
  \bibfield  {author} {\bibinfo {author} {\bibfnamefont {L.}~\bibnamefont
  {Huang}}, \bibinfo {author} {\bibfnamefont {Q.}~\bibnamefont {Chen}},
  \bibinfo {author} {\bibfnamefont {Y.-C.}\ \bibnamefont {Lai}}, \ and\
  \bibinfo {author} {\bibfnamefont {L.~M.}\ \bibnamefont {Pecora}},\ }\bibfield
   {title} {\enquote {\bibinfo {title} {Generic behavior of master-stability
  functions in coupled nonlinear dynamical systems},}\ }\href {\doibase
  10.1103/PhysRevE.80.036204} {\bibfield  {journal} {\bibinfo  {journal} {Phys.
  Rev. E}\ }\textbf {\bibinfo {volume} {80}},\ \bibinfo {pages} {036204}
  (\bibinfo {year} {2009})}\BibitemShut {NoStop}%
\bibitem [{\citenamefont {Watanabe}\ and\ \citenamefont
  {Strogatz}(1993)}]{WS:1993}%
  \BibitemOpen
  \bibfield  {author} {\bibinfo {author} {\bibfnamefont {S.}~\bibnamefont
  {Watanabe}}\ and\ \bibinfo {author} {\bibfnamefont {S.~H.}\ \bibnamefont
  {Strogatz}},\ }\bibfield  {title} {\enquote {\bibinfo {title} {Integrability
  of a globally coupled oscillator array},}\ }\href {\doibase
  10.1103/PhysRevLett.70.2391} {\bibfield  {journal} {\bibinfo  {journal}
  {Phys. Rev. Lett.}\ }\textbf {\bibinfo {volume} {70}},\ \bibinfo {pages}
  {2391} (\bibinfo {year} {1993})}\BibitemShut {NoStop}%
\bibitem [{\citenamefont {Ott}\ and\ \citenamefont {Antonsen}(2008)}]{OA:2008}%
  \BibitemOpen
  \bibfield  {author} {\bibinfo {author} {\bibfnamefont {E.}~\bibnamefont
  {Ott}}\ and\ \bibinfo {author} {\bibfnamefont {T.~M.}\ \bibnamefont
  {Antonsen}},\ }\bibfield  {title} {\enquote {\bibinfo {title} {Low
  dimensional behavior of large systems of globally coupled oscillators},}\
  }\href {\doibase 10.1063/1.2930766} {\bibfield  {journal} {\bibinfo
  {journal} {Chaos}\ }\textbf {\bibinfo {volume} {18}},\ \bibinfo {pages}
  {037113} (\bibinfo {year} {2008})}\BibitemShut {NoStop}%
\bibitem [{\citenamefont {Williams}\ \emph
  {et~al.}(2013{\natexlab{a}})\citenamefont {Williams}, \citenamefont {Murphy},
  \citenamefont {Roy}, \citenamefont {Sorrentino}, \citenamefont {Dahms},\ and\
  \citenamefont {Sch\"oll}}]{WMRSDS:2013}%
  \BibitemOpen
  \bibfield  {author} {\bibinfo {author} {\bibfnamefont {C.~R.~S.}\
  \bibnamefont {Williams}}, \bibinfo {author} {\bibfnamefont {T.~E.}\
  \bibnamefont {Murphy}}, \bibinfo {author} {\bibfnamefont {R.}~\bibnamefont
  {Roy}}, \bibinfo {author} {\bibfnamefont {F.}~\bibnamefont {Sorrentino}},
  \bibinfo {author} {\bibfnamefont {T.}~\bibnamefont {Dahms}}, \ and\ \bibinfo
  {author} {\bibfnamefont {E.}~\bibnamefont {Sch\"oll}},\ }\bibfield  {title}
  {\enquote {\bibinfo {title} {Experimental observations of group synchrony in
  a system of chaotic optoelectronic oscillators},}\ }\href {\doibase
  10.1103/PhysRevLett.110.064104} {\bibfield  {journal} {\bibinfo  {journal}
  {Phys. Rev. Lett.}\ }\textbf {\bibinfo {volume} {110}},\ \bibinfo {pages}
  {064104} (\bibinfo {year} {2013}{\natexlab{a}})}\BibitemShut {NoStop}%
\bibitem [{\citenamefont {Williams}\ \emph
  {et~al.}(2013{\natexlab{b}})\citenamefont {Williams}, \citenamefont
  {Sorrentino}, \citenamefont {Murphy},\ and\ \citenamefont {Roy}}]{WSMR:2013}%
  \BibitemOpen
  \bibfield  {author} {\bibinfo {author} {\bibfnamefont {C.}~\bibnamefont
  {Williams}}, \bibinfo {author} {\bibfnamefont {F.}~\bibnamefont
  {Sorrentino}}, \bibinfo {author} {\bibfnamefont {T.~E.}\ \bibnamefont
  {Murphy}}, \ and\ \bibinfo {author} {\bibfnamefont {R.}~\bibnamefont {Roy}},\
  }\bibfield  {title} {\enquote {\bibinfo {title} {Synchronization states and
  multistability in a ring of periodic oscillators: Experimentally variable
  coupling delays},}\ }\href@noop {} {\bibfield  {journal} {\bibinfo  {journal}
  {Chaos}\ }\textbf {\bibinfo {volume} {23}},\ \bibinfo {pages} {143117}
  (\bibinfo {year} {2013}{\natexlab{b}})}\BibitemShut {NoStop}%
\bibitem [{\citenamefont {Hart}\ \emph {et~al.}(2015)\citenamefont {Hart},
  \citenamefont {Pade}, \citenamefont {Pereira}, \citenamefont {Murphy},\ and\
  \citenamefont {Roy}}]{HPPMR:2015}%
  \BibitemOpen
  \bibfield  {author} {\bibinfo {author} {\bibfnamefont {J.~D.}\ \bibnamefont
  {Hart}}, \bibinfo {author} {\bibfnamefont {J.~P.}\ \bibnamefont {Pade}},
  \bibinfo {author} {\bibfnamefont {T.}~\bibnamefont {Pereira}}, \bibinfo
  {author} {\bibfnamefont {T.~E.}\ \bibnamefont {Murphy}}, \ and\ \bibinfo
  {author} {\bibfnamefont {R.}~\bibnamefont {Roy}},\ }\bibfield  {title}
  {\enquote {\bibinfo {title} {Adding connections can hinder network
  synchronization of time-delayed oscillators},}\ }\href {\doibase
  10.1103/PhysRevE.92.022804} {\bibfield  {journal} {\bibinfo  {journal} {Phys.
  Rev. E}\ }\textbf {\bibinfo {volume} {92}},\ \bibinfo {pages} {022804}
  (\bibinfo {year} {2015})}\BibitemShut {NoStop}%
\bibitem [{\citenamefont {Arenas}\ \emph {et~al.}(2008)\citenamefont {Arenas},
  \citenamefont {D{\'\i}az-Guilera}, \citenamefont {Kurths}, \citenamefont
  {Moreno},\ and\ \citenamefont {Zhou}}]{ADKMZ:2008}%
  \BibitemOpen
  \bibfield  {author} {\bibinfo {author} {\bibfnamefont {A.}~\bibnamefont
  {Arenas}}, \bibinfo {author} {\bibfnamefont {A.}~\bibnamefont
  {D{\'\i}az-Guilera}}, \bibinfo {author} {\bibfnamefont {J.}~\bibnamefont
  {Kurths}}, \bibinfo {author} {\bibfnamefont {Y.}~\bibnamefont {Moreno}}, \
  and\ \bibinfo {author} {\bibfnamefont {C.~S.}\ \bibnamefont {Zhou}},\
  }\bibfield  {title} {\enquote {\bibinfo {title} {Synchronization in complex
  networks},}\ }\href {\doibase https://doi.org/10.1016/j.physrep.2008.09.002}
  {\bibfield  {journal} {\bibinfo  {journal} {Phy. Rep.}\ }\textbf {\bibinfo
  {volume} {469}},\ \bibinfo {pages} {93 } (\bibinfo {year}
  {2008})}\BibitemShut {NoStop}%
\bibitem [{\citenamefont {Barahona}\ and\ \citenamefont
  {Pecora}(2002)}]{BP:2002}%
  \BibitemOpen
  \bibfield  {author} {\bibinfo {author} {\bibfnamefont {M.}~\bibnamefont
  {Barahona}}\ and\ \bibinfo {author} {\bibfnamefont {L.~M.}\ \bibnamefont
  {Pecora}},\ }\bibfield  {title} {\enquote {\bibinfo {title} {Synchronization
  in small-world systems},}\ }\href@noop {} {\bibfield  {journal} {\bibinfo
  {journal} {Phys. Rev. Lett.}\ }\textbf {\bibinfo {volume} {89}},\ \bibinfo
  {pages} {054101} (\bibinfo {year} {2002})}\BibitemShut {NoStop}%
\bibitem [{\citenamefont {Nishikawa}\ \emph {et~al.}(2003)\citenamefont
  {Nishikawa}, \citenamefont {Motter}, \citenamefont {Lai},\ and\ \citenamefont
  {Hoppensteadt}}]{NMLH:2003}%
  \BibitemOpen
  \bibfield  {author} {\bibinfo {author} {\bibfnamefont {T.}~\bibnamefont
  {Nishikawa}}, \bibinfo {author} {\bibfnamefont {A.~E.}\ \bibnamefont
  {Motter}}, \bibinfo {author} {\bibfnamefont {Y.-C.}\ \bibnamefont {Lai}}, \
  and\ \bibinfo {author} {\bibfnamefont {F.~C.}\ \bibnamefont {Hoppensteadt}},\
  }\bibfield  {title} {\enquote {\bibinfo {title} {Heterogeneity in oscillator
  networks: Are smaller worlds easier to synchronize?}}\ }\href {\doibase
  10.1103/PhysRevLett.91.014101} {\bibfield  {journal} {\bibinfo  {journal}
  {Phys. Rev. Lett.}\ }\textbf {\bibinfo {volume} {91}},\ \bibinfo {pages}
  {014101} (\bibinfo {year} {2003})}\BibitemShut {NoStop}%
\bibitem [{\citenamefont {Wang}\ \emph {et~al.}(2007)\citenamefont {Wang},
  \citenamefont {Lai},\ and\ \citenamefont {Lai}}]{WLL:2007}%
  \BibitemOpen
  \bibfield  {author} {\bibinfo {author} {\bibfnamefont {X.~G.}\ \bibnamefont
  {Wang}}, \bibinfo {author} {\bibfnamefont {Y.-C.}\ \bibnamefont {Lai}}, \
  and\ \bibinfo {author} {\bibfnamefont {C.~H.}\ \bibnamefont {Lai}},\
  }\bibfield  {title} {\enquote {\bibinfo {title} {Enhancing synchronization
  based on complex gradient networks},}\ }\href {\doibase
  10.1103/PhysRevE.75.056205} {\bibfield  {journal} {\bibinfo  {journal} {Phys.
  Rev. E}\ }\textbf {\bibinfo {volume} {75}},\ \bibinfo {pages} {056205}
  (\bibinfo {year} {2007})}\BibitemShut {NoStop}%
\bibitem [{\citenamefont {Ricci}\ \emph {et~al.}(2012)\citenamefont {Ricci},
  \citenamefont {Tonelli}, \citenamefont {Huang},\ and\ \citenamefont
  {Lai}}]{RTHL:2012}%
  \BibitemOpen
  \bibfield  {author} {\bibinfo {author} {\bibfnamefont {F.}~\bibnamefont
  {Ricci}}, \bibinfo {author} {\bibfnamefont {R.}~\bibnamefont {Tonelli}},
  \bibinfo {author} {\bibfnamefont {L.}~\bibnamefont {Huang}}, \ and\ \bibinfo
  {author} {\bibfnamefont {Y.-C.}\ \bibnamefont {Lai}},\ }\bibfield  {title}
  {\enquote {\bibinfo {title} {Onset of chaotic phase synchronization in
  complex networks of coupled heterogeneous oscillators},}\ }\href {\doibase
  10.1103/PhysRevE.86.027201} {\bibfield  {journal} {\bibinfo  {journal} {Phys.
  Rev. E}\ }\textbf {\bibinfo {volume} {86}},\ \bibinfo {pages} {027201}
  (\bibinfo {year} {2012})}\BibitemShut {NoStop}%
\bibitem [{\citenamefont {Pecora}\ \emph {et~al.}(2014)\citenamefont {Pecora},
  \citenamefont {Sorrentino}, \citenamefont {Hagerstrom}, \citenamefont
  {Murphy},\ and\ \citenamefont {Roy}}]{PSHMR:2014}%
  \BibitemOpen
  \bibfield  {author} {\bibinfo {author} {\bibfnamefont {L.~M.}\ \bibnamefont
  {Pecora}}, \bibinfo {author} {\bibfnamefont {F.}~\bibnamefont {Sorrentino}},
  \bibinfo {author} {\bibfnamefont {A.~M.}\ \bibnamefont {Hagerstrom}},
  \bibinfo {author} {\bibfnamefont {T.~E.}\ \bibnamefont {Murphy}}, \ and\
  \bibinfo {author} {\bibfnamefont {R.}~\bibnamefont {Roy}},\ }\bibfield
  {title} {\enquote {\bibinfo {title} {Cluster synchronization and isolated
  desynchronization in complex networks with symmetries},}\ }\href@noop {}
  {\bibfield  {journal} {\bibinfo  {journal} {Nat. Commun.}\ }\textbf {\bibinfo
  {volume} {5}},\ \bibinfo {pages} {4079} (\bibinfo {year} {2014})}\BibitemShut
  {NoStop}%
\bibitem [{\citenamefont {Sorrentino}\ \emph {et~al.}(2016)\citenamefont
  {Sorrentino}, \citenamefont {Pecora}, \citenamefont {Hagerstrom},
  \citenamefont {Murphy},\ and\ \citenamefont {Roy}}]{SPHMR:2016}%
  \BibitemOpen
  \bibfield  {author} {\bibinfo {author} {\bibfnamefont {F.}~\bibnamefont
  {Sorrentino}}, \bibinfo {author} {\bibfnamefont {L.~M.}\ \bibnamefont
  {Pecora}}, \bibinfo {author} {\bibfnamefont {A.~M.}\ \bibnamefont
  {Hagerstrom}}, \bibinfo {author} {\bibfnamefont {T.~E.}\ \bibnamefont
  {Murphy}}, \ and\ \bibinfo {author} {\bibfnamefont {R.}~\bibnamefont {Roy}},\
  }\bibfield  {title} {\enquote {\bibinfo {title} {Cluster synchronization and
  isolated desynchronization in complex networks with symmetries},}\
  }\href@noop {} {\bibfield  {journal} {\bibinfo  {journal} {Sci. Adv.}\
  }\textbf {\bibinfo {volume} {2}},\ \bibinfo {pages} {e1501737} (\bibinfo
  {year} {2016})}\BibitemShut {NoStop}%
\bibitem [{\citenamefont {G\'omez-Garde\~nes}\ \emph
  {et~al.}(2011)\citenamefont {G\'omez-Garde\~nes}, \citenamefont {G\'omez},
  \citenamefont {Arenas},\ and\ \citenamefont {Moreno}}]{GGAM:2011}%
  \BibitemOpen
  \bibfield  {author} {\bibinfo {author} {\bibfnamefont {J.}~\bibnamefont
  {G\'omez-Garde\~nes}}, \bibinfo {author} {\bibfnamefont {S.}~\bibnamefont
  {G\'omez}}, \bibinfo {author} {\bibfnamefont {A.}~\bibnamefont {Arenas}}, \
  and\ \bibinfo {author} {\bibfnamefont {Y.}~\bibnamefont {Moreno}},\
  }\bibfield  {title} {\enquote {\bibinfo {title} {Explosive synchronization
  transitions in scale-free networks},}\ }\href {\doibase
  10.1103/PhysRevLett.106.128701} {\bibfield  {journal} {\bibinfo  {journal}
  {Phys. Rev. Lett.}\ }\textbf {\bibinfo {volume} {106}},\ \bibinfo {pages}
  {128701} (\bibinfo {year} {2011})}\BibitemShut {NoStop}%
\bibitem [{\citenamefont {Zou}\ \emph {et~al.}(2014)\citenamefont {Zou},
  \citenamefont {Pereira}, \citenamefont {Small}, \citenamefont {Liu},\ and\
  \citenamefont {Kurths}}]{ZPSLK:2014}%
  \BibitemOpen
  \bibfield  {author} {\bibinfo {author} {\bibfnamefont {Y.}~\bibnamefont
  {Zou}}, \bibinfo {author} {\bibfnamefont {T.}~\bibnamefont {Pereira}},
  \bibinfo {author} {\bibfnamefont {M.}~\bibnamefont {Small}}, \bibinfo
  {author} {\bibfnamefont {Z.}~\bibnamefont {Liu}}, \ and\ \bibinfo {author}
  {\bibfnamefont {J.}~\bibnamefont {Kurths}},\ }\bibfield  {title} {\enquote
  {\bibinfo {title} {Basin of attraction determines hysteresis in explosive
  synchronization},}\ }\href {\doibase 10.1103/PhysRevLett.112.114102}
  {\bibfield  {journal} {\bibinfo  {journal} {Phys. Rev. Lett.}\ }\textbf
  {\bibinfo {volume} {112}},\ \bibinfo {pages} {114102} (\bibinfo {year}
  {2014})}\BibitemShut {NoStop}%
\bibitem [{\citenamefont {Boccaletti}\ \emph {et~al.}(2016)\citenamefont
  {Boccaletti}, \citenamefont {Almendral}, \citenamefont {Guan}, \citenamefont
  {Leyva}, \citenamefont {Liu}, \citenamefont {Sendi{\~n}a-Nadal},
  \citenamefont {Wang},\ and\ \citenamefont {Zou}}]{BAGLLSWZ:2016}%
  \BibitemOpen
  \bibfield  {author} {\bibinfo {author} {\bibfnamefont {S.}~\bibnamefont
  {Boccaletti}}, \bibinfo {author} {\bibfnamefont {J.~A.}\ \bibnamefont
  {Almendral}}, \bibinfo {author} {\bibfnamefont {S.}~\bibnamefont {Guan}},
  \bibinfo {author} {\bibfnamefont {I.}~\bibnamefont {Leyva}}, \bibinfo
  {author} {\bibfnamefont {Z.}~\bibnamefont {Liu}}, \bibinfo {author}
  {\bibfnamefont {I.}~\bibnamefont {Sendi{\~n}a-Nadal}}, \bibinfo {author}
  {\bibfnamefont {Z.}~\bibnamefont {Wang}}, \ and\ \bibinfo {author}
  {\bibfnamefont {Y.}~\bibnamefont {Zou}},\ }\bibfield  {title} {\enquote
  {\bibinfo {title} {Explosive transitions in complex networks' structure and
  dynamics: Percolation and synchronization},}\ }\href@noop {} {\bibfield
  {journal} {\bibinfo  {journal} {Phys. Rep.}\ }\textbf {\bibinfo {volume}
  {660}},\ \bibinfo {pages} {1} (\bibinfo {year} {2016})}\BibitemShut {NoStop}%
\bibitem [{\citenamefont {Lipton}\ \emph {et~al.}(2015)\citenamefont {Lipton},
  \citenamefont {Berkowitz},\ and\ \citenamefont {Elkan}}]{LBE:2015}%
  \BibitemOpen
  \bibfield  {author} {\bibinfo {author} {\bibfnamefont {Z.~C.}\ \bibnamefont
  {Lipton}}, \bibinfo {author} {\bibfnamefont {J.}~\bibnamefont {Berkowitz}}, \
  and\ \bibinfo {author} {\bibfnamefont {C.}~\bibnamefont {Elkan}},\ }\bibfield
   {title} {\enquote {\bibinfo {title} {A critical review of recurrent neural
  networks for sequence learning},}\ }\href@noop {} {\bibfield  {journal}
  {\bibinfo  {journal} {arXiv:1506.00019}\ } (\bibinfo {year}
  {2015})}\BibitemShut {NoStop}%
\bibitem [{\citenamefont {Haynes}\ \emph {et~al.}(2015)\citenamefont {Haynes},
  \citenamefont {Soriano}, \citenamefont {Rosin}, \citenamefont {Fischer},\
  and\ \citenamefont {Gauthier}}]{HSRFG:2015}%
  \BibitemOpen
  \bibfield  {author} {\bibinfo {author} {\bibfnamefont {N.~D.}\ \bibnamefont
  {Haynes}}, \bibinfo {author} {\bibfnamefont {M.~C.}\ \bibnamefont {Soriano}},
  \bibinfo {author} {\bibfnamefont {D.~P.}\ \bibnamefont {Rosin}}, \bibinfo
  {author} {\bibfnamefont {I.}~\bibnamefont {Fischer}}, \ and\ \bibinfo
  {author} {\bibfnamefont {D.~J.}\ \bibnamefont {Gauthier}},\ }\bibfield
  {title} {\enquote {\bibinfo {title} {Reservoir computing with a single
  time-delay autonomous {Boolean} node},}\ }\href {\doibase
  10.1103/PhysRevE.91.020801} {\bibfield  {journal} {\bibinfo  {journal} {Phys.
  Rev. E}\ }\textbf {\bibinfo {volume} {91}},\ \bibinfo {pages} {020801}
  (\bibinfo {year} {2015})}\BibitemShut {NoStop}%
\bibitem [{\citenamefont {Larger}\ \emph {et~al.}(2017)\citenamefont {Larger},
  \citenamefont {Bayl\'on-Fuentes}, \citenamefont {Martinenghi}, \citenamefont
  {Udaltsov}, \citenamefont {Chembo},\ and\ \citenamefont
  {Jacquot}}]{LBMUCJ:2017}%
  \BibitemOpen
  \bibfield  {author} {\bibinfo {author} {\bibfnamefont {L.}~\bibnamefont
  {Larger}}, \bibinfo {author} {\bibfnamefont {A.}~\bibnamefont
  {Bayl\'on-Fuentes}}, \bibinfo {author} {\bibfnamefont {R.}~\bibnamefont
  {Martinenghi}}, \bibinfo {author} {\bibfnamefont {V.~S.}\ \bibnamefont
  {Udaltsov}}, \bibinfo {author} {\bibfnamefont {Y.~K.}\ \bibnamefont
  {Chembo}}, \ and\ \bibinfo {author} {\bibfnamefont {M.}~\bibnamefont
  {Jacquot}},\ }\bibfield  {title} {\enquote {\bibinfo {title} {High-speed
  photonic reservoir computing using a time-delay-based architecture: Million
  words per second classification},}\ }\href {\doibase
  10.1103/PhysRevX.7.011015} {\bibfield  {journal} {\bibinfo  {journal} {Phys.
  Rev. X}\ }\textbf {\bibinfo {volume} {7}},\ \bibinfo {pages} {011015}
  (\bibinfo {year} {2017})}\BibitemShut {NoStop}%
\bibitem [{\citenamefont {Pathak}\ \emph {et~al.}(2017)\citenamefont {Pathak},
  \citenamefont {Lu}, \citenamefont {Hunt}, \citenamefont {Girvan},\ and\
  \citenamefont {Ott}}]{PLHGO:2017}%
  \BibitemOpen
  \bibfield  {author} {\bibinfo {author} {\bibfnamefont {J.}~\bibnamefont
  {Pathak}}, \bibinfo {author} {\bibfnamefont {Z.}~\bibnamefont {Lu}}, \bibinfo
  {author} {\bibfnamefont {B.~R.}\ \bibnamefont {Hunt}}, \bibinfo {author}
  {\bibfnamefont {M.}~\bibnamefont {Girvan}}, \ and\ \bibinfo {author}
  {\bibfnamefont {E.}~\bibnamefont {Ott}},\ }\bibfield  {title} {\enquote
  {\bibinfo {title} {Using machine learning to replicate chaotic attractors and
  calculate lyapunov exponents from data},}\ }\href {\doibase
  10.1063/1.5010300} {\bibfield  {journal} {\bibinfo  {journal} {Chaos}\
  }\textbf {\bibinfo {volume} {27}},\ \bibinfo {pages} {121102} (\bibinfo
  {year} {2017})}\BibitemShut {NoStop}%
\bibitem [{\citenamefont {Lu}\ \emph {et~al.}(2017)\citenamefont {Lu},
  \citenamefont {Pathak}, \citenamefont {Hunt}, \citenamefont {Girvan},
  \citenamefont {Brockett},\ and\ \citenamefont {Ott}}]{LPHGBO:2017}%
  \BibitemOpen
  \bibfield  {author} {\bibinfo {author} {\bibfnamefont {Z.}~\bibnamefont
  {Lu}}, \bibinfo {author} {\bibfnamefont {J.}~\bibnamefont {Pathak}}, \bibinfo
  {author} {\bibfnamefont {B.}~\bibnamefont {Hunt}}, \bibinfo {author}
  {\bibfnamefont {M.}~\bibnamefont {Girvan}}, \bibinfo {author} {\bibfnamefont
  {R.}~\bibnamefont {Brockett}}, \ and\ \bibinfo {author} {\bibfnamefont
  {E.}~\bibnamefont {Ott}},\ }\bibfield  {title} {\enquote {\bibinfo {title}
  {Reservoir observers: Model-free inference of unmeasured variables in chaotic
  systems},}\ }\href {\doibase 10.1063/1.4979665} {\bibfield  {journal}
  {\bibinfo  {journal} {Chaos}\ }\textbf {\bibinfo {volume} {27}},\ \bibinfo
  {pages} {041102} (\bibinfo {year} {2017})}\BibitemShut {NoStop}%
\bibitem [{\citenamefont {Duriez}\ \emph {et~al.}(2017)\citenamefont {Duriez},
  \citenamefont {Brunton},\ and\ \citenamefont {Noack}}]{DBN:book}%
  \BibitemOpen
  \bibfield  {author} {\bibinfo {author} {\bibfnamefont {T.}~\bibnamefont
  {Duriez}}, \bibinfo {author} {\bibfnamefont {S.~L.}\ \bibnamefont {Brunton}},
  \ and\ \bibinfo {author} {\bibfnamefont {B.~R.}\ \bibnamefont {Noack}},\
  }\href@noop {} {\emph {\bibinfo {title} {Machine Learning Control-Taming
  Nonlinear Dynamics and Turbulence}}}\ (\bibinfo  {publisher} {Springer},\
  \bibinfo {year} {2017})\BibitemShut {NoStop}%
\bibitem [{\citenamefont {Lu}\ \emph {et~al.}(2018)\citenamefont {Lu},
  \citenamefont {Hunt},\ and\ \citenamefont {Ott}}]{LHO:2018}%
  \BibitemOpen
  \bibfield  {author} {\bibinfo {author} {\bibfnamefont {Z.}~\bibnamefont
  {Lu}}, \bibinfo {author} {\bibfnamefont {B.~R.}\ \bibnamefont {Hunt}}, \ and\
  \bibinfo {author} {\bibfnamefont {E.}~\bibnamefont {Ott}},\ }\bibfield
  {title} {\enquote {\bibinfo {title} {Attractor reconstruction by machine
  learning},}\ }\href@noop {} {\bibfield  {journal} {\bibinfo  {journal}
  {Chaos}\ }\textbf {\bibinfo {volume} {28}},\ \bibinfo {pages} {061104}
  (\bibinfo {year} {2018})}\BibitemShut {NoStop}%
\bibitem [{\citenamefont {Pathak}\ \emph
  {et~al.}(2018{\natexlab{a}})\citenamefont {Pathak}, \citenamefont {Wilner},
  \citenamefont {Fussell}, \citenamefont {Chandra}, \citenamefont {Hunt},
  \citenamefont {Girvan}, \citenamefont {Lu},\ and\ \citenamefont
  {Ott}}]{PWFCHGO:2018}%
  \BibitemOpen
  \bibfield  {author} {\bibinfo {author} {\bibfnamefont {J.}~\bibnamefont
  {Pathak}}, \bibinfo {author} {\bibfnamefont {A.}~\bibnamefont {Wilner}},
  \bibinfo {author} {\bibfnamefont {R.}~\bibnamefont {Fussell}}, \bibinfo
  {author} {\bibfnamefont {S.}~\bibnamefont {Chandra}}, \bibinfo {author}
  {\bibfnamefont {B.}~\bibnamefont {Hunt}}, \bibinfo {author} {\bibfnamefont
  {M.}~\bibnamefont {Girvan}}, \bibinfo {author} {\bibfnamefont
  {Z.}~\bibnamefont {Lu}}, \ and\ \bibinfo {author} {\bibfnamefont
  {E.}~\bibnamefont {Ott}},\ }\bibfield  {title} {\enquote {\bibinfo {title}
  {Hybrid forecasting of chaotic processes: Using machine learning in
  conjunction with a knowledge-based model},}\ }\href@noop {} {\bibfield
  {journal} {\bibinfo  {journal} {Chaos}\ }\textbf {\bibinfo {volume} {28}},\
  \bibinfo {pages} {041101} (\bibinfo {year} {2018}{\natexlab{a}})}\BibitemShut
  {NoStop}%
\bibitem [{\citenamefont {Pathak}\ \emph
  {et~al.}(2018{\natexlab{b}})\citenamefont {Pathak}, \citenamefont {Hunt},
  \citenamefont {Girvan}, \citenamefont {Lu},\ and\ \citenamefont
  {Ott}}]{PHGLO:2018}%
  \BibitemOpen
  \bibfield  {author} {\bibinfo {author} {\bibfnamefont {J.}~\bibnamefont
  {Pathak}}, \bibinfo {author} {\bibfnamefont {B.}~\bibnamefont {Hunt}},
  \bibinfo {author} {\bibfnamefont {M.}~\bibnamefont {Girvan}}, \bibinfo
  {author} {\bibfnamefont {Z.}~\bibnamefont {Lu}}, \ and\ \bibinfo {author}
  {\bibfnamefont {E.}~\bibnamefont {Ott}},\ }\bibfield  {title} {\enquote
  {\bibinfo {title} {Model-free prediction of large spatiotemporally chaotic
  systems from data: A reservoir computing approach},}\ }\href {\doibase
  10.1103/PhysRevLett.120.024102} {\bibfield  {journal} {\bibinfo  {journal}
  {Phys. Rev. Lett.}\ }\textbf {\bibinfo {volume} {120}},\ \bibinfo {pages}
  {024102} (\bibinfo {year} {2018}{\natexlab{b}})}\BibitemShut {NoStop}%
\bibitem [{\citenamefont {Carroll}(2018)}]{Carroll:2018}%
  \BibitemOpen
  \bibfield  {author} {\bibinfo {author} {\bibfnamefont {T.~L.}\ \bibnamefont
  {Carroll}},\ }\bibfield  {title} {\enquote {\bibinfo {title} {Using reservoir
  computers to distinguish chaotic signals},}\ }\href {\doibase
  10.1103/PhysRevE.98.052209} {\bibfield  {journal} {\bibinfo  {journal} {Phys.
  Rev. E}\ }\textbf {\bibinfo {volume} {98}},\ \bibinfo {pages} {052209}
  (\bibinfo {year} {2018})}\BibitemShut {NoStop}%
\bibitem [{\citenamefont {Nakai}\ and\ \citenamefont {Saiki}(2018)}]{NS:2018}%
  \BibitemOpen
  \bibfield  {author} {\bibinfo {author} {\bibfnamefont {K.}~\bibnamefont
  {Nakai}}\ and\ \bibinfo {author} {\bibfnamefont {Y.}~\bibnamefont {Saiki}},\
  }\bibfield  {title} {\enquote {\bibinfo {title} {Machine-learning inference
  of fluid variables from data using reservoir computing},}\ }\href {\doibase
  10.1103/PhysRevE.98.023111} {\bibfield  {journal} {\bibinfo  {journal} {Phys.
  Rev. E}\ }\textbf {\bibinfo {volume} {98}},\ \bibinfo {pages} {023111}
  (\bibinfo {year} {2018})}\BibitemShut {NoStop}%
\bibitem [{\citenamefont {Zimmermann}\ and\ \citenamefont
  {Parlitz}(2018)}]{ZP:2018}%
  \BibitemOpen
  \bibfield  {author} {\bibinfo {author} {\bibfnamefont {R.~S.}\ \bibnamefont
  {Zimmermann}}\ and\ \bibinfo {author} {\bibfnamefont {U.}~\bibnamefont
  {Parlitz}},\ }\bibfield  {title} {\enquote {\bibinfo {title} {Observing
  spatio-temporal dynamics of excitable media using reservoir computing},}\
  }\href {\doibase 10.1063/1.5022276} {\bibfield  {journal} {\bibinfo
  {journal} {Chaos}\ }\textbf {\bibinfo {volume} {28}},\ \bibinfo {pages}
  {043118} (\bibinfo {year} {2018})}\BibitemShut {NoStop}%
\bibitem [{\citenamefont {Weng}\ \emph {et~al.}(2019)\citenamefont {Weng},
  \citenamefont {Yang}, \citenamefont {Gu}, \citenamefont {Zhang},\ and\
  \citenamefont {Small}}]{WYGZS:2019}%
  \BibitemOpen
  \bibfield  {author} {\bibinfo {author} {\bibfnamefont {T.}~\bibnamefont
  {Weng}}, \bibinfo {author} {\bibfnamefont {H.}~\bibnamefont {Yang}}, \bibinfo
  {author} {\bibfnamefont {C.}~\bibnamefont {Gu}}, \bibinfo {author}
  {\bibfnamefont {J.}~\bibnamefont {Zhang}}, \ and\ \bibinfo {author}
  {\bibfnamefont {M.}~\bibnamefont {Small}},\ }\bibfield  {title} {\enquote
  {\bibinfo {title} {Synchronization of chaotic systems and their
  machine-learning models},}\ }\href {\doibase 10.1103/PhysRevE.99.042203}
  {\bibfield  {journal} {\bibinfo  {journal} {Phys. Rev. E}\ }\textbf {\bibinfo
  {volume} {99}},\ \bibinfo {pages} {042203} (\bibinfo {year}
  {2019})}\BibitemShut {NoStop}%
\bibitem [{\citenamefont {Griffith}\ \emph {et~al.}(2019)\citenamefont
  {Griffith}, \citenamefont {Pomerance},\ and\ \citenamefont
  {Gauthier}}]{GPG:2019}%
  \BibitemOpen
  \bibfield  {author} {\bibinfo {author} {\bibfnamefont {A.}~\bibnamefont
  {Griffith}}, \bibinfo {author} {\bibfnamefont {A.}~\bibnamefont {Pomerance}},
  \ and\ \bibinfo {author} {\bibfnamefont {D.~J.}\ \bibnamefont {Gauthier}},\
  }\bibfield  {title} {\enquote {\bibinfo {title} {Forecasting chaotic systems
  with very low connectivity reservoir computers},}\ }\href@noop {} {\bibfield
  {journal} {\bibinfo  {journal} {Chaos}\ }\textbf {\bibinfo {volume} {29}},\
  \bibinfo {pages} {123108} (\bibinfo {year} {2019})}\BibitemShut {NoStop}%
\bibitem [{\citenamefont {Jiang}\ and\ \citenamefont {Lai}(2019)}]{JL:2019}%
  \BibitemOpen
  \bibfield  {author} {\bibinfo {author} {\bibfnamefont {J.}~\bibnamefont
  {Jiang}}\ and\ \bibinfo {author} {\bibfnamefont {Y.-C.}\ \bibnamefont
  {Lai}},\ }\bibfield  {title} {\enquote {\bibinfo {title} {Model-free
  prediction of spatiotemporal dynamical systems with recurrent neural
  networks: Role of network spectral radius},}\ }\href {\doibase
  10.1103/PhysRevResearch.1.033056} {\bibfield  {journal} {\bibinfo  {journal}
  {Phys. Rev. Research}\ }\textbf {\bibinfo {volume} {1}},\ \bibinfo {pages}
  {033056} (\bibinfo {year} {2019})}\BibitemShut {NoStop}%
\bibitem [{\citenamefont {Vlachas}\ \emph {et~al.}(2019)\citenamefont
  {Vlachas}, \citenamefont {Pathak}, \citenamefont {Hunt}, \citenamefont
  {Sapsis}, \citenamefont {Girvan}, \citenamefont {Ott},\ and\ \citenamefont
  {Koumoutsakos}}]{VPHSGOK:2019}%
  \BibitemOpen
  \bibfield  {author} {\bibinfo {author} {\bibfnamefont {P.~R.}\ \bibnamefont
  {Vlachas}}, \bibinfo {author} {\bibfnamefont {J.}~\bibnamefont {Pathak}},
  \bibinfo {author} {\bibfnamefont {B.~R.}\ \bibnamefont {Hunt}}, \bibinfo
  {author} {\bibfnamefont {T.~P.}\ \bibnamefont {Sapsis}}, \bibinfo {author}
  {\bibfnamefont {M.}~\bibnamefont {Girvan}}, \bibinfo {author} {\bibfnamefont
  {E.}~\bibnamefont {Ott}}, \ and\ \bibinfo {author} {\bibfnamefont
  {P.}~\bibnamefont {Koumoutsakos}},\ }\bibfield  {title} {\enquote {\bibinfo
  {title} {Forecasting of spatio-temporal chaotic dynamics with recurrent
  neural networks: A comparative study of reservoir computing and
  backpropagation algorithms},}\ }\href@noop {} {\bibfield  {journal} {\bibinfo
   {journal} {arXiv preprint arXiv:1910.05266}\ } (\bibinfo {year}
  {2019})}\BibitemShut {NoStop}%
\bibitem [{\citenamefont {Fan}\ \emph {et~al.}(2020)\citenamefont {Fan},
  \citenamefont {Jiang}, \citenamefont {Zhang}, \citenamefont {Wang},\ and\
  \citenamefont {Lai}}]{FJZWL:2020}%
  \BibitemOpen
  \bibfield  {author} {\bibinfo {author} {\bibfnamefont {H.}~\bibnamefont
  {Fan}}, \bibinfo {author} {\bibfnamefont {J.}~\bibnamefont {Jiang}}, \bibinfo
  {author} {\bibfnamefont {C.}~\bibnamefont {Zhang}}, \bibinfo {author}
  {\bibfnamefont {X.}~\bibnamefont {Wang}}, \ and\ \bibinfo {author}
  {\bibfnamefont {Y.-C.}\ \bibnamefont {Lai}},\ }\bibfield  {title} {\enquote
  {\bibinfo {title} {Long-term prediction of chaotic systems with machine
  learning},}\ }\href {\doibase 10.1103/PhysRevResearch.2.012080} {\bibfield
  {journal} {\bibinfo  {journal} {Phys. Rev. Research}\ }\textbf {\bibinfo
  {volume} {2}},\ \bibinfo {pages} {012080} (\bibinfo {year}
  {2020})}\BibitemShut {NoStop}%
\bibitem [{\citenamefont {Zhang}\ \emph {et~al.}(2020)\citenamefont {Zhang},
  \citenamefont {Jiang}, \citenamefont {Qu},\ and\ \citenamefont
  {Lai}}]{ZJQL:2020}%
  \BibitemOpen
  \bibfield  {author} {\bibinfo {author} {\bibfnamefont {C.}~\bibnamefont
  {Zhang}}, \bibinfo {author} {\bibfnamefont {J.}~\bibnamefont {Jiang}},
  \bibinfo {author} {\bibfnamefont {S.-X.}\ \bibnamefont {Qu}}, \ and\ \bibinfo
  {author} {\bibfnamefont {Y.-C.}\ \bibnamefont {Lai}},\ }\bibfield  {title}
  {\enquote {\bibinfo {title} {Predicting phase and sensing phase coherence in
  chaotic systems with machine learning},}\ }\href@noop {} {\bibfield
  {journal} {\bibinfo  {journal} {Chaos}\ }\textbf {\bibinfo {volume} {30}},\
  \bibinfo {pages} {083114} (\bibinfo {year} {2020})}\BibitemShut {NoStop}%
\bibitem [{\citenamefont {Guo}\ \emph {et~al.}(2021)\citenamefont {Guo},
  \citenamefont {Zhang}, \citenamefont {Wang}, \citenamefont {Fan},
  \citenamefont {Xiao},\ and\ \citenamefont {Wang}}]{GYL:2021}%
  \BibitemOpen
  \bibfield  {author} {\bibinfo {author} {\bibfnamefont {Y.}~\bibnamefont
  {Guo}}, \bibinfo {author} {\bibfnamefont {H.}~\bibnamefont {Zhang}}, \bibinfo
  {author} {\bibfnamefont {L.}~\bibnamefont {Wang}}, \bibinfo {author}
  {\bibfnamefont {H.}~\bibnamefont {Fan}}, \bibinfo {author} {\bibfnamefont
  {J.}~\bibnamefont {Xiao}}, \ and\ \bibinfo {author} {\bibfnamefont
  {X.}~\bibnamefont {Wang}},\ }\bibfield  {title} {\enquote {\bibinfo {title}
  {Transfer learning of chaotic systems},}\ }\href {\doibase 10.1063/5.0033870}
  {\bibfield  {journal} {\bibinfo  {journal} {Chaos}\ }\textbf {\bibinfo
  {volume} {31}},\ \bibinfo {pages} {011104} (\bibinfo {year}
  {2021})}\BibitemShut {NoStop}%
\bibitem [{\citenamefont {Cestnik}\ and\ \citenamefont {Abel}(2019)}]{CA:2019}%
  \BibitemOpen
  \bibfield  {author} {\bibinfo {author} {\bibfnamefont {R.}~\bibnamefont
  {Cestnik}}\ and\ \bibinfo {author} {\bibfnamefont {M.}~\bibnamefont {Abel}},\
  }\bibfield  {title} {\enquote {\bibinfo {title} {Inferring the dynamics of
  oscillatory systems using recurrent neural networks},}\ }\href {\doibase
  10.1063/1.5096918} {\bibfield  {journal} {\bibinfo  {journal} {Chaos}\
  }\textbf {\bibinfo {volume} {29}},\ \bibinfo {pages} {063128} (\bibinfo
  {year} {2019})}\BibitemShut {NoStop}%
\bibitem [{\citenamefont {Kong}\ \emph {et~al.}(2021)\citenamefont {Kong},
  \citenamefont {Fan}, \citenamefont {Grebogi},\ and\ \citenamefont
  {Lai}}]{KFGL:2021}%
  \BibitemOpen
  \bibfield  {author} {\bibinfo {author} {\bibfnamefont {L.-W.}\ \bibnamefont
  {Kong}}, \bibinfo {author} {\bibfnamefont {H.-W.}\ \bibnamefont {Fan}},
  \bibinfo {author} {\bibfnamefont {C.}~\bibnamefont {Grebogi}}, \ and\
  \bibinfo {author} {\bibfnamefont {Y.-C.}\ \bibnamefont {Lai}},\ }\bibfield
  {title} {\enquote {\bibinfo {title} {Machine learning prediction of critical
  transition and system collapse},}\ }\href {\doibase
  10.1103/PhysRevResearch.3.013090} {\bibfield  {journal} {\bibinfo  {journal}
  {Phys. Rev. Research}\ }\textbf {\bibinfo {volume} {3}},\ \bibinfo {pages}
  {013090} (\bibinfo {year} {2021})}\BibitemShut {NoStop}%
\bibitem [{\citenamefont {Wang}\ \emph
  {et~al.}(2011{\natexlab{a}})\citenamefont {Wang}, \citenamefont {Yang},
  \citenamefont {Lai}, \citenamefont {Kovanis},\ and\ \citenamefont
  {Grebogi}}]{WYLKG:2011}%
  \BibitemOpen
  \bibfield  {author} {\bibinfo {author} {\bibfnamefont {W.-X.}\ \bibnamefont
  {Wang}}, \bibinfo {author} {\bibfnamefont {R.}~\bibnamefont {Yang}}, \bibinfo
  {author} {\bibfnamefont {Y.-C.}\ \bibnamefont {Lai}}, \bibinfo {author}
  {\bibfnamefont {V.}~\bibnamefont {Kovanis}}, \ and\ \bibinfo {author}
  {\bibfnamefont {C.}~\bibnamefont {Grebogi}},\ }\bibfield  {title} {\enquote
  {\bibinfo {title} {Predicting catastrophes in nonlinear dynamical systems by
  compressive sensing},}\ }\href@noop {} {\bibfield  {journal} {\bibinfo
  {journal} {Phys. Rev. Lett.}\ }\textbf {\bibinfo {volume} {106}},\ \bibinfo
  {pages} {154101} (\bibinfo {year} {2011}{\natexlab{a}})}\BibitemShut
  {NoStop}%
\bibitem [{\citenamefont {Wang}\ \emph
  {et~al.}(2011{\natexlab{b}})\citenamefont {Wang}, \citenamefont {Yang},
  \citenamefont {Lai}, \citenamefont {Kovanis},\ and\ \citenamefont
  {Harrison}}]{WYLKH:2011}%
  \BibitemOpen
  \bibfield  {author} {\bibinfo {author} {\bibfnamefont {W.-X.}\ \bibnamefont
  {Wang}}, \bibinfo {author} {\bibfnamefont {R.}~\bibnamefont {Yang}}, \bibinfo
  {author} {\bibfnamefont {Y.-C.}\ \bibnamefont {Lai}}, \bibinfo {author}
  {\bibfnamefont {V.}~\bibnamefont {Kovanis}}, \ and\ \bibinfo {author}
  {\bibfnamefont {M.~A.~F.}\ \bibnamefont {Harrison}},\ }\bibfield  {title}
  {\enquote {\bibinfo {title} {Time-series-based prediction of complex
  oscillator networks via compressive sensing},}\ }\href@noop {} {\bibfield
  {journal} {\bibinfo  {journal} {EPL (Europhys. Lett.)}\ }\textbf {\bibinfo
  {volume} {94}},\ \bibinfo {pages} {48006} (\bibinfo {year}
  {2011}{\natexlab{b}})}\BibitemShut {NoStop}%
\bibitem [{\citenamefont {Wang}\ \emph {et~al.}(2016)\citenamefont {Wang},
  \citenamefont {Lai},\ and\ \citenamefont {Grebogi}}]{WLG:2016}%
  \BibitemOpen
  \bibfield  {author} {\bibinfo {author} {\bibfnamefont {W.}~\bibnamefont
  {Wang}}, \bibinfo {author} {\bibfnamefont {Y.-C.}\ \bibnamefont {Lai}}, \
  and\ \bibinfo {author} {\bibfnamefont {C.}~\bibnamefont {Grebogi}},\
  }\bibfield  {title} {\enquote {\bibinfo {title} {Data based identification
  and prediction of nonlinear and complex dynamical systems},}\ }\href@noop {}
  {\bibfield  {journal} {\bibinfo  {journal} {Phys. Rep.}\ }\textbf {\bibinfo
  {volume} {644}},\ \bibinfo {pages} {1} (\bibinfo {year} {2016})}\BibitemShut
  {NoStop}%
\bibitem [{\citenamefont {Su}\ \emph {et~al.}(2012)\citenamefont {Su},
  \citenamefont {Ni}, \citenamefont {Wang},\ and\ \citenamefont
  {Lai}}]{SNWL:2012}%
  \BibitemOpen
  \bibfield  {author} {\bibinfo {author} {\bibfnamefont {R.-Q.}\ \bibnamefont
  {Su}}, \bibinfo {author} {\bibfnamefont {X.}~\bibnamefont {Ni}}, \bibinfo
  {author} {\bibfnamefont {W.-X.}\ \bibnamefont {Wang}}, \ and\ \bibinfo
  {author} {\bibfnamefont {Y.-C.}\ \bibnamefont {Lai}},\ }\bibfield  {title}
  {\enquote {\bibinfo {title} {Forecasting synchronizability of complex
  networks from data},}\ }\href {\doibase 10.1103/PhysRevE.85.056220}
  {\bibfield  {journal} {\bibinfo  {journal} {Phys. Rev. E}\ }\textbf {\bibinfo
  {volume} {85}},\ \bibinfo {pages} {056220} (\bibinfo {year}
  {2012})}\BibitemShut {NoStop}%
\bibitem [{\citenamefont {Lorenz}(1963)}]{Lorenz:1963}%
  \BibitemOpen
  \bibfield  {author} {\bibinfo {author} {\bibfnamefont {E.~N.}\ \bibnamefont
  {Lorenz}},\ }\bibfield  {title} {\enquote {\bibinfo {title} {Deterministic
  nonperiodic flow},}\ }\href@noop {} {\bibfield  {journal} {\bibinfo
  {journal} {J. Atmos. Sci.}\ }\textbf {\bibinfo {volume} {20}},\ \bibinfo
  {pages} {130} (\bibinfo {year} {1963})}\BibitemShut {NoStop}%
\bibitem [{\citenamefont {McCann}\ and\ \citenamefont
  {Yodzis}(1994)}]{MY:1994}%
  \BibitemOpen
  \bibfield  {author} {\bibinfo {author} {\bibfnamefont {K.}~\bibnamefont
  {McCann}}\ and\ \bibinfo {author} {\bibfnamefont {P.}~\bibnamefont
  {Yodzis}},\ }\bibfield  {title} {\enquote {\bibinfo {title} {Nonlinear
  dynamics and population disappearances},}\ }\href@noop {} {\bibfield
  {journal} {\bibinfo  {journal} {Ame. Naturalist}\ }\textbf {\bibinfo {volume}
  {144}},\ \bibinfo {pages} {873} (\bibinfo {year} {1994})}\BibitemShut
  {NoStop}%
\end{thebibliography}

%
\end{document}